\RequirePackage{lineno}
\documentclass[aps,floatfix,prc,twocolumn,amsmath,amssymb,superscriptaddress]{revtex4}

\usepackage{soul}
\usepackage{graphicx}	
\usepackage{xspace}     
\usepackage{amsmath}
\usepackage{hyperref}
\usepackage{color}
\usepackage{multirow}
\usepackage{booktabs,array}
\RequirePackage{lineno}
\newcommand{\pT}{$p_{\rm T}$\xspace}
\newcommand{\pTavg}{$\langle p_{\rm T} \rangle$\xspace}

\newcommand{\Sref}[1]{Section~\ref{#1}}

\newcommand{\Au}{$\mathrm{Au}+\mathrm{Au}$\xspace}
\newcommand{\Pb}{$\mathrm{Pb}+\mathrm{Pb}$\xspace}

\newcommand{\sNN}{$\sqrt{s_{\mathrm{NN}}}$}

\def	\sNN 	{\sqrt{s_{NN}}} 

\begin{document}

\title{Constraining the particle production mechanism in Au + Au collisions
  at $\sqrt{s_{NN}} =$ 7.7, 27, and 200 GeV using a multiphase transport model}

\author{Abhirikshma Nandi} 
\affiliation{Indian Institute of Technology Guwahati, Assam, India-781039.}

\author{Lokesh Kumar} 
\author{Natasha Sharma} 
\affiliation{Department of Physics, Panjab University, Chandigarh, India-160014.}

\begin{abstract} 
We study the production of pions, kaons, and (anti)protons using a
multiphase transport (AMPT) model in Au + Au collisions at $\sqrt{s_{NN}}=$ 7.7,
27, and 200 GeV.  
We present the centrality and energy dependence of various bulk
observables such as invariant yields as a function of transverse
momentum $p_T$, particle yields $dN/dy$, average transverse momentum
$\langle p_T \rangle$,  and various particle ratios, and compare them with experimental
data. Both default and string melting (SM) versions of the AMPT model
are used with three different sets of initial conditions. We observe
that neither the default nor the SM version of the model could
consistently describe the centrality dependence of all observables at
the above energies with any one set of initial conditions.  
The energy dependence behavior of the experimental observables for
0--5\% central collisions is in general better described by 
the default AMPT model using the modified HIJING parameters for Lund
string fragmentation and 3 mb parton scattering cross section. In
addition, the kaon production as well as the $K/\pi$ ratio at 7.7 GeV
are under predicted by
the AMPT model. 

\end{abstract}

\maketitle

\section{Introduction}

Relativistic collisions of heavy ions make it possible to subject nuclear matter to the
extreme energy densities required for a possible deconfinement of quarks and gluons. A dense
matter with partonic degrees of freedom, often called the quark-gluon plasma (QGP), is
expected to form in the initial moments after the
collision~\cite{Adams:2005dq,Adcox:2004mh,Arsene:2004fa,Back:2004je}. Exploring
the quantum chromodynamics (QCD) phase  diagram to understand the
properties of quark matter is one of the most important goals of
high-energy 
heavy-ion
experiments~\cite{Laermann:2003cv,Rajagopal:2000wf,Stephanov:2007fk}. Comparing
the results obtained from theoretical models with the experimental
data helps in understanding the space-time evolution of QGP and many
of its other properties. The QCD phase diagram is usually plotted as
temperature ($T$) versus 
baryon chemical potential ($\mu_{\mathrm{B}}$). Assuming a thermalized system is reached in heavy-ion
collisions, both $T$ and 
$\mu_B$ can be varied by changing the collision
energy~\cite{Cleymans:1999st,Becattini:2005xt,Andronic:2005yp}. To this
end, the Beam 
Energy Scan (BES) program
at the BNL Relativistic Heavy Ion Collider (RHIC) completed its first
phase of operation in 2010 and
2011~\cite{Kumar:2015yia,Kumar:2011de,Abelev:2009bw,Mohanty:2009vb,Aggarwal:2010cw,Kumar:2012fb,Kumar:2013cqa,Adamczyk:2017iwn}.  
The measurements of the bulk properties of identified hadrons using the BES data were recently published~\cite{Adamczyk:2017iwn}. 
The measurements from the STAR experiment cover the $\mu_{\mathrm{B}}$ interval from 20 to 450 MeV. This is also
believed to be the region in which the transition from hadronic matter
to QGP takes place~\cite{Adamczyk:2013dal, Adamczyk:2014ipa,
  Adamczyk:2014mzf,Adamczyk:2017nof,Adamczyk:2015ukd,Adamczyk:2013gv,Adamczyk:2017nxg}.  

In this paper, we have studied \Au
collisions at $\sNN$ = 7.7, 27, and 200 GeV using a multi phase transport (AMPT) model and compared bulk properties such as transverse momentum \pT spectra, multiplicity densities $dN/dy$, average transverse momentum \pTavg and particle ratios with the experimental data. 
For this study we have used three different sets of parameters for
both the default and string melting (SM) versions of the AMPT
model. 
It may be noted that comparisons of the elliptic flow $v_2$ from AMPT
with the experimental data from RHIC BES energies have been done
previously~\cite{Adamczyk:2012ku,Adamczyk:2015fum}. This is, however,
the first time that the hadron production has been compared in such detail
with the experimental data at these energies. The purpose of this work
is to use already tuned parameters for the AMPT model and to see which
set of parameters or physics describes the data best and consistently
at these energies and different centralities. 
Thus, we take one parameter set at a time and study its
  energy and centrality dependence. In total, we consider three different
  sets, as motivated by previous studies~\cite{Lin:2004en,Zhu:2015voa,He:2017tla,Xu:2011fi}. 
This would help in understanding the particle production in heavy-ion collisions. 
It may, however, happen that different parameters
  work at different centralities and/or energies. 

The paper is organized as follows. In \Sref{Sec:Model} we give a brief
description of the AMPT model and its parameters. In
\Sref{Sec:Spectra} we present the comparison of transverse momentum
spectra between models and experimental data. In \Sref{Sec:Yield} and
\Sref{Sec:ptaverage} we study the centrality dependence of particle
yields and average transverse momenta respectively and compare the
results with experimental data. The centrality and energy dependence
of various particle ratios are discussed in \Sref{Sec:Ratios} and
\Sref{Sec:EdepRat} respectively. We summarize in
\Sref{Sec:Conclusions}.

\section{The AMPT Model}\label{Sec:Model}
In this section, we give a short description of the AMPT model and its parameters. The AMPT model was developed to give a coherent description of the dynamics of relativistic heavy-ion collisions~\cite{Lin:2004en} and has been used extensively to study them at various energies and environments. 
It is a hybrid transport model and has four main components: the initial conditions, partonic interactions, hadronization, and hadronic interactions~\cite{Lin:2004en}. Initial conditions are obtained from the Heavy Ion Jet Interaction Generator (HIJING) model~\cite{Wang:1991hta}. Hard minijet partons are produced perturbatively if the momentum transfer is more than a threshold ($p_{0}$ = 2 GeV/c) and soft strings are produced otherwise. Depending on the version of AMPT model used, default or string melting, the soft strings are either retained or are completely converted to partons. 

Zhangs's parton cascade (ZPC)~\cite{Zhang:1997ej} is used for partonic interactions. 
The differential scattering cross-section is given by\begin{equation}
\frac{d\sigma}{dt} \approx \frac{9\pi\alpha_{s}^{2}}{2(t-\mu^{2})^{2}},
\end{equation}
where $\sigma$ is the parton-parton scattering cross section, $t$ is the standard Mandelstam variable for four-momentum transfer, $\alpha_{s}$ is the strong coupling constant, and $\mu$ is the Debye screening mass in partonic matter. 

In the default model, only the minijet gluons take part in the ZPC and the energy stored in the excited strings is only released after hadrons are formed. For the default model, after the partons stop interacting, they combine with their parent strings. Hadronization of these strings take place using the Lund string fragmentation model~\cite{Andersson:1983jt,Andersson:1983ia}. The longitudinal momentum of the hadrons generated is given by the Lund string fragmentation function $f(z) \propto z^{-1}(1-z)^{a} \exp(-b m_{T}^{2}/z)$, $z$ being the light-cone momentum fraction of the hadron of transverse mass $m_{T}$ with respect to the fragmenting string. 
The average squared transverse momentum $\langle p_{T}^{2} \rangle$ of the produced particles is proportional to the string tension $\kappa$, i.e., the energy stored per unit length of a string, and depends on the Lund string fragmentation parameters as \begin{equation}
\label{eq:2}
\kappa \propto \langle p_{T}^{2} \rangle = \frac{1}{b(2 + a)}.
\end{equation} 

\begin{figure*}[htb]
\begin{center}
  \includegraphics[width = 0.9\textwidth]{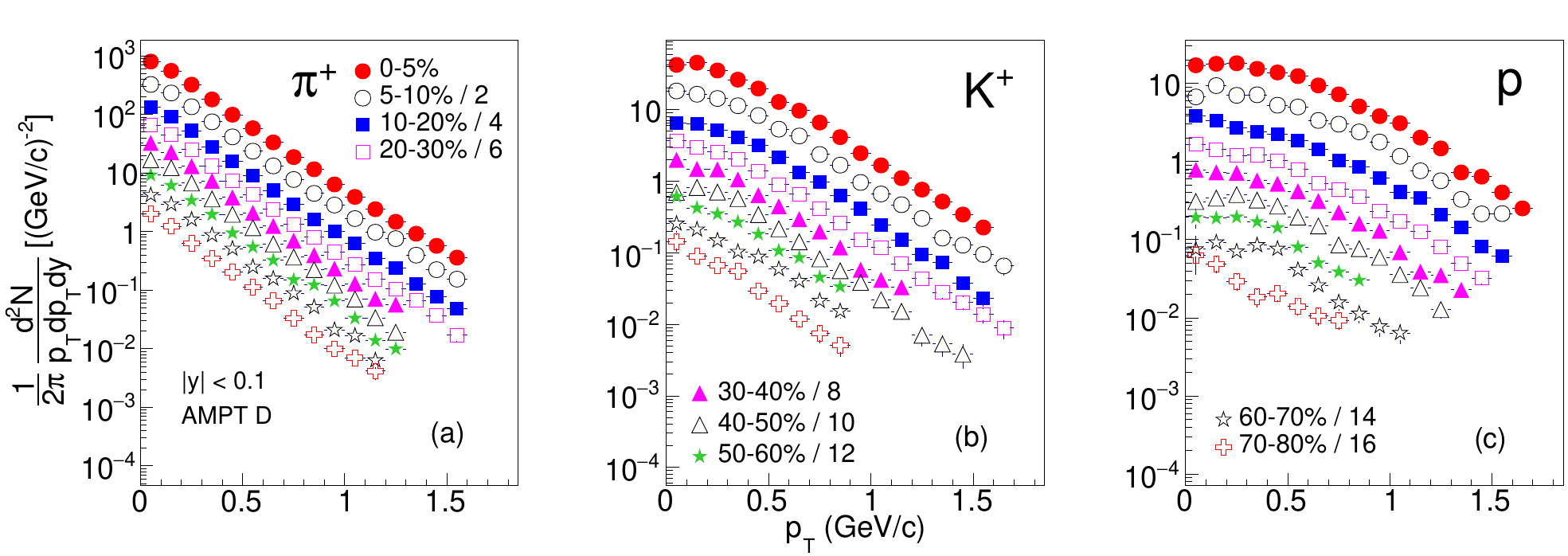}
  \includegraphics[width = 0.9\textwidth]{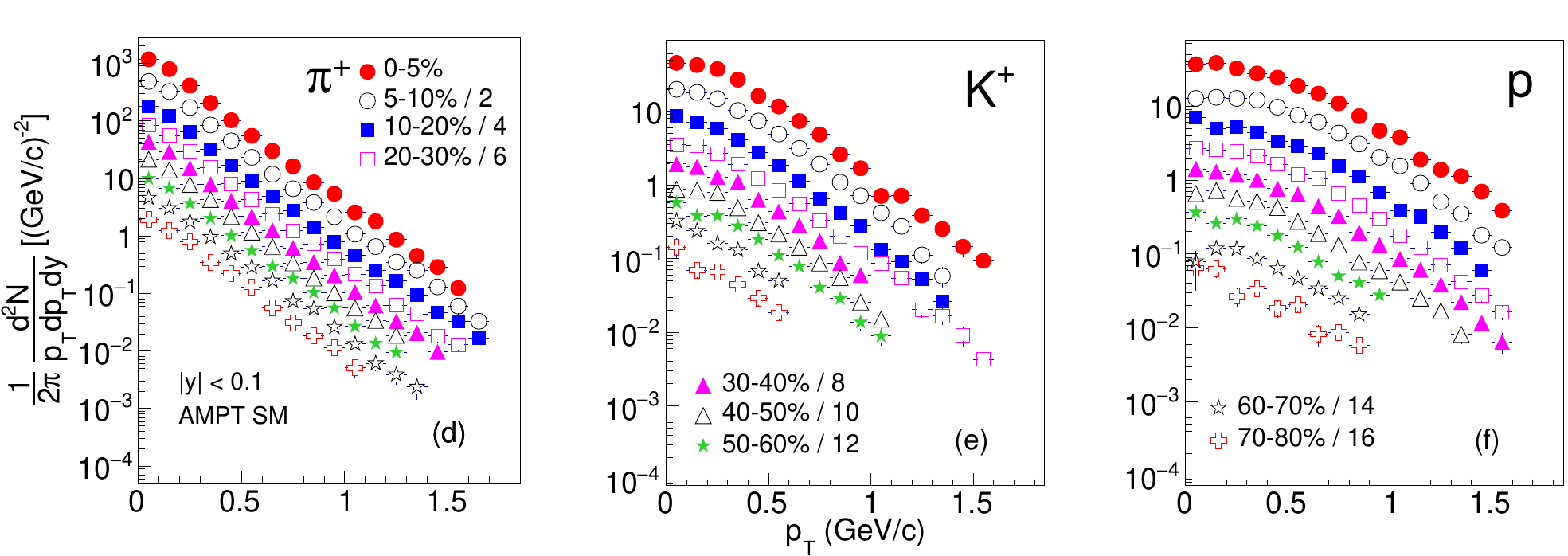}
  \caption{Midrapidity ($|y| < 0.1$) transverse momentum spectra of
    $\pi^{+}$, $K^{+}$, $p$ in \Au collisions at $\sNN$ = 27 GeV at
    different centralities using set B (HIJING default a = 0.5, b =
    0.9 GeV$^{-2}$, and $\sigma$ = 1.5mb) parameters in default [(a)--(c)] and string melting [(d)--(f)] versions of the AMPT model. Spectra for centralities other than 0-5\% are scaled for clarity as shown in figure.}
  \label{fig:1}
\end{center}
\end{figure*}
In the string melting version, quarks and antiquarks of all flavors take part in the ZPC, and hadronization takes place via a quark
coalescence model in which the nearest partons are combined to form
mesons and baryons. The dynamics of the hadronic matter is described
by a relativistic transport (ART) model which includes meson-meson,
meson-baryon, baryon-baryon, elastic, and inelastic
scatterings~\cite{Li:1995pra}, but the hadronic effects are less important than in the default model. The parton density in ZPC for the SM
version is quite high as all HIJING strings are converted to
partons. As a result the SM version was found to reasonably fit the
elliptic flow at RHIC~\cite{Lin:2004en}.
It was observed that the SM version could not describe the hadron production, specifically the proton rapidity distributions or the $p_T$ spectra of hadrons~\cite{Lin:2004en,Zhu:2015voa,He:2017tla}. However, it was shown in a recent work that an improved quark coalescence could improve the description of hadron production, but more work was needed to be done to improve the baryon and strangeness sector~\cite{He:2017tla}.

\begin{table}[htb]
  \centering
	\caption{Used values of parameters in Lund string fragmentation and parton scattering cross sections for the three sets of AMPT data.}
	\label{tab:lsff}\vspace{0.1in}
	\begin{tabular}{|c|c|c|c|c|c|}
	\hline	
Set &	Cross section $\sigma$ & $a$ & $b$ (GeV$^{-2}$) & $\alpha_{s}$ & $\mu$ (fm$^{-1}$) \\ \hline
Set A  &	3 mb   & 0.55 & 0.15 & 0.33 & 2.265 \\
Set B  &	1.5 mb & 0.5 & 0.9 & 0.33 & 3.2 \\
Set C  &	10 mb  & 2.2 & 0.5 & 0.47 & 1.8 \\
	\hline
	\end{tabular}
\end{table}
We have chosen the three tuned parameter sets as given in Table~\ref{tab:lsff}. Some of these parameters might depend on the collision conditions, e.g., the Lund fragmentation parameters a and b are expected to depend on the collision centrality. The string tension $\kappa$, which depends on a and b, might also be environment dependent due to close packing of strings~\cite{Fischer:2016zzs}. In the model, the screening mass $\mu$ is a parameter for changing the parton scattering cross sections but it is generated by medium effects. So, a medium dependent screening mass would produce results truer to real physics. Instead of using different parameters for different collision conditions, three sets of constant parameters are used for this study. 
These have been chosen by taking guidance from earlier studies as
detailed below.
 The parton scattering cross section is given as $\sigma \approx 9 \pi \alpha_{s}^{2}/(2 \mu^{2})$. Thus, the value of  $\sigma$ depends on a given combination of $\alpha_s$ and $\mu$.
It has been observed that the multiplicity is not very sensitive to the parton scattering cross section $\sigma$~\cite{Xu:2011fi}, but $\sigma$ seems to affect the elliptic flow such that a larger parton scattering cross section leads to large elliptic flows~\cite{Xu:2011fi}.

It has been observed that the default HIJING values for the Lund string fragmentation parameters ($a=0.5$ and $b=0.9$ GeV$^{-2}$) in set B were able to describe the $pp$ data when used in the AMPT default model but underestimated the charged particle yield in central \Pb collisions at the top Super Proton Synchrotron (SPS) energy~\cite{Lin:2014tya,Zhang:1999bd,Lin:2000cx}.
For Pb+Pb collisions at Large Hadron Collider (LHC) energies, the AMPT SM model with default HIJING values for the Lund string fragmentation parameters  ($a=0.5$ and $b=0.9$ GeV$^{-2}$) in set B was able to reproduce the yield and elliptic flow of charged particles but underestimated the \pT spectrum except at low \pT~\cite{Lin:2014tya,Xu:2011fi}.

From Eq. (\ref{eq:2}) it is clear that parameters $a$ and $b$ determine the \pT distribution of the particles. For larger $a$ and $b$ there will be a smaller average square transverse momentum that will produce a steeper \pT spectrum (with large slope), while their smaller values will lead to a flatter distribution.
 It has been reported that the values of $a=2.2$ and $b=0.5$ GeV$^{-2}$
produce larger multiplicity density as compared to other values of $a$ and $b$~\cite{Xu:2011fi}. 
Thus, the modified values of  $a=2.2$ and $b=0.5$ GeV$^{-2}$ (set C) were used to fit the charged particle yield in Pb+Pb collisions at SPS~\cite{Lin:2014tya, Lin:2000cx}. For heavy-ion collisions at RHIC energies, the default AMPT model with these parameters was found to reasonably fit the rapidity and pseudorapidity density and the \pT spectra but underestimate the elliptic flow~\cite{Lin:2014tya, Lin:2000cx}. On using the AMPT SM with same parameters, the elliptic flow and two-pion Hanbury Brown and Twiss correlations (HBT) were reproduced but the charged particle yield was overestimated while the slopes of the \pT spectra were underestimated~\cite{Lin:2014tya, Lin:2004en}.

In order to simultaneously fit the rapidity density, \pT spectrum, and elliptic flow of pions and kaons at low \pT in Au+Au collisions at RHIC energies, the AMPT SM model was used with modified Lund string fragmentation parameters $a=0.55$ and $b=0.15$ GeV$^{-2}$ in set A~\cite{Lin:2014tya}.

Thus we observe that each of these sets satisfactorily describes the heavy-ion data at different energies from various experiments.
 The availability of centrality dependent results at the RHIC for a vast range of energies allows us to test the validity of the given parameters at these conditions.
  We have generated AMPT events for Au+Au collisions at three energies, viz., the lowest RHIC energy (7.7 GeV), an intermediate energy (27 GeV), and the top RHIC energy of 200 GeV. 
  The model version ampt-v1.26t7-v2.26t7 is used for this study.
 The events are generated
  using both string melting and default versions of the AMPT. For each
  of these versions, we have used the three sets of parameters listed in
  Table~\ref{tab:lsff} to generate the events. 
  About 20000 events are used for the analysis at each energy, for each set, and for each of the two versions of the model.
  The centrality selection is done in the same way as in the experimental data~\cite{Adamczyk:2017iwn}. Thus, the AMPT data are divided into nine centrality classes: 0--5\%, 5--10\%, 10--20\%, 20--30\%, 30--40\%, 40--50\%, 50--60\%, 60--70\%, and 70--80\%.

\section{Results}
We present the mid-rapidity ($|y| < 0.1$) transverse momentum \pT spectra, particle yields $dN/dy$, average transverse momentum \pTavg and ratios of identified particles $\pi^{\pm}$, $K^{\pm}$, p, and $\bar{p}$ at $\sNN$ = 7.7, 27, and 200 GeV.
The results are obtained for both AMPT SM and default versions at each energy and using three different sets of parameters listed in Table~\ref{tab:lsff}. The simulated results are compared with corresponding results from the STAR experiment. 
\subsection{Transverse momentum spectra}
\begin{figure*}[!]
\begin{center}
  \includegraphics[width = \textwidth]{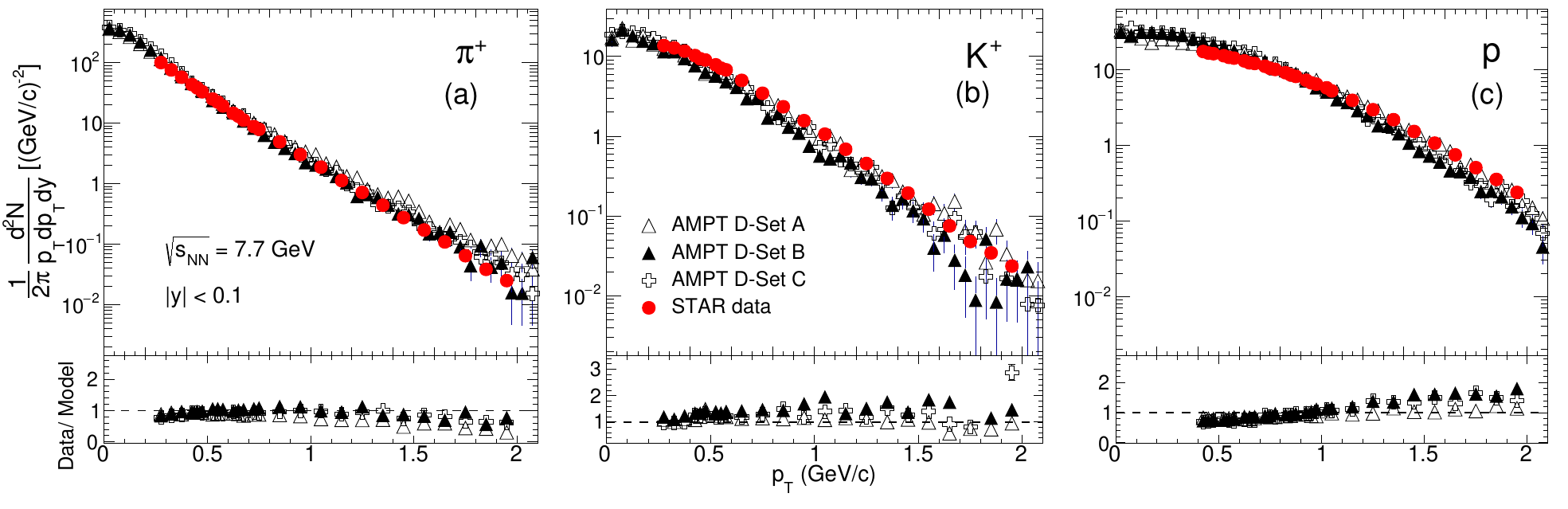}
  \includegraphics[width = \textwidth]{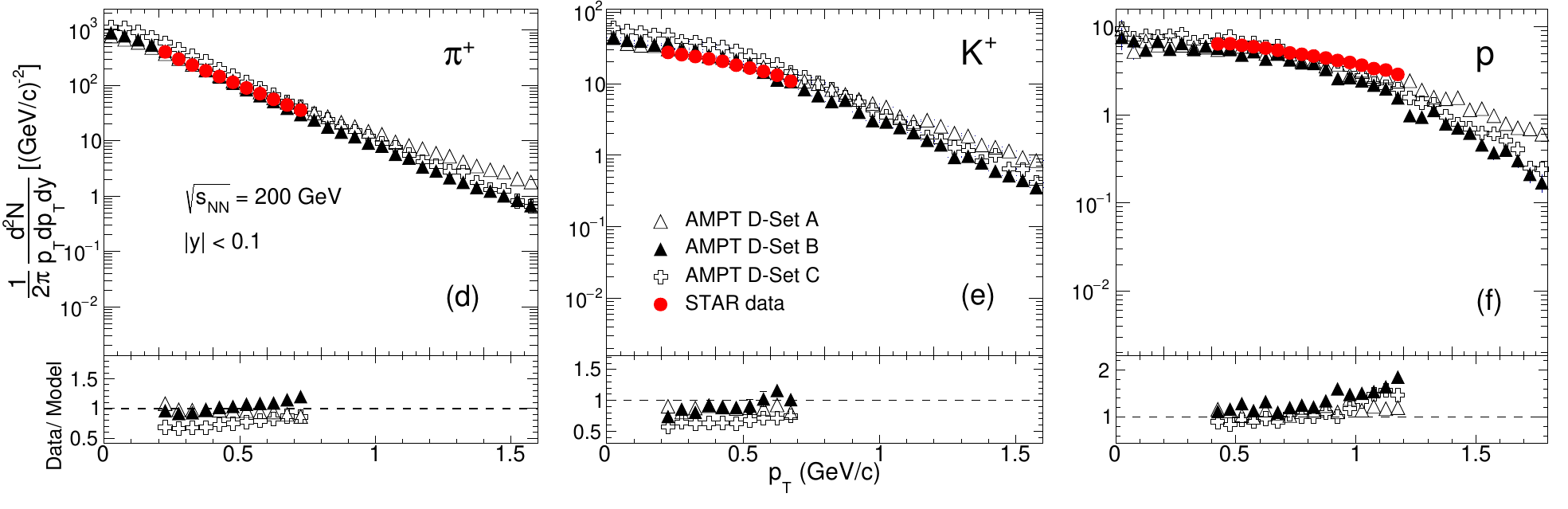}
  \includegraphics[width = \textwidth]{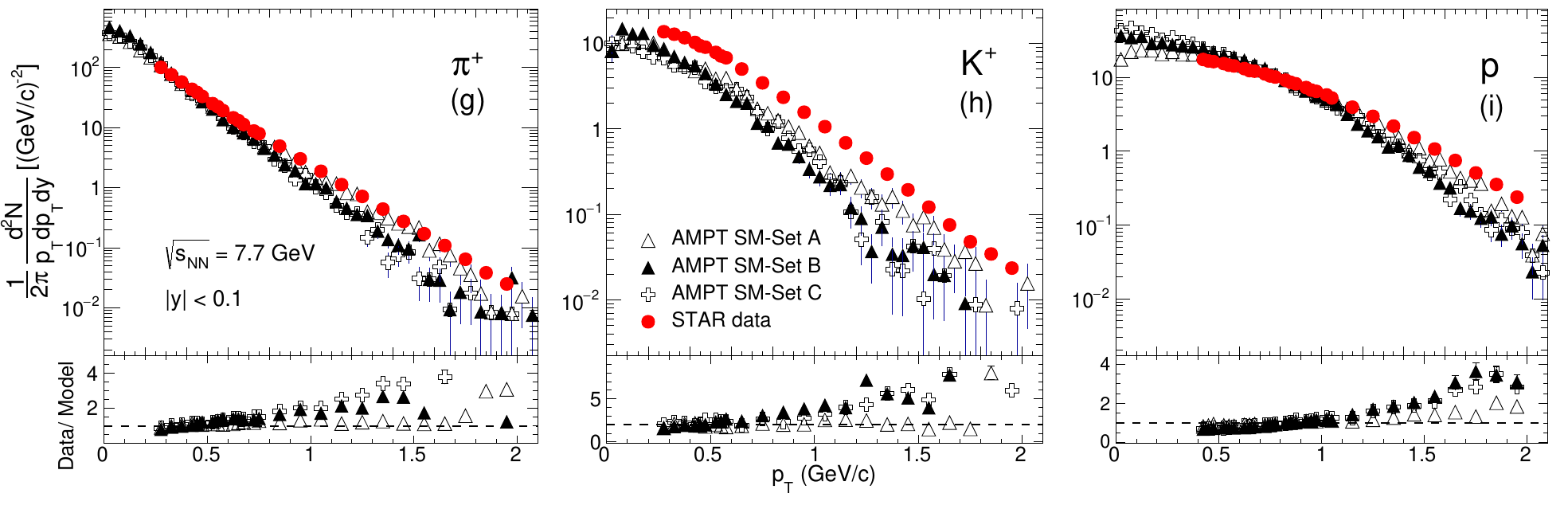}
  \includegraphics[width = \textwidth]{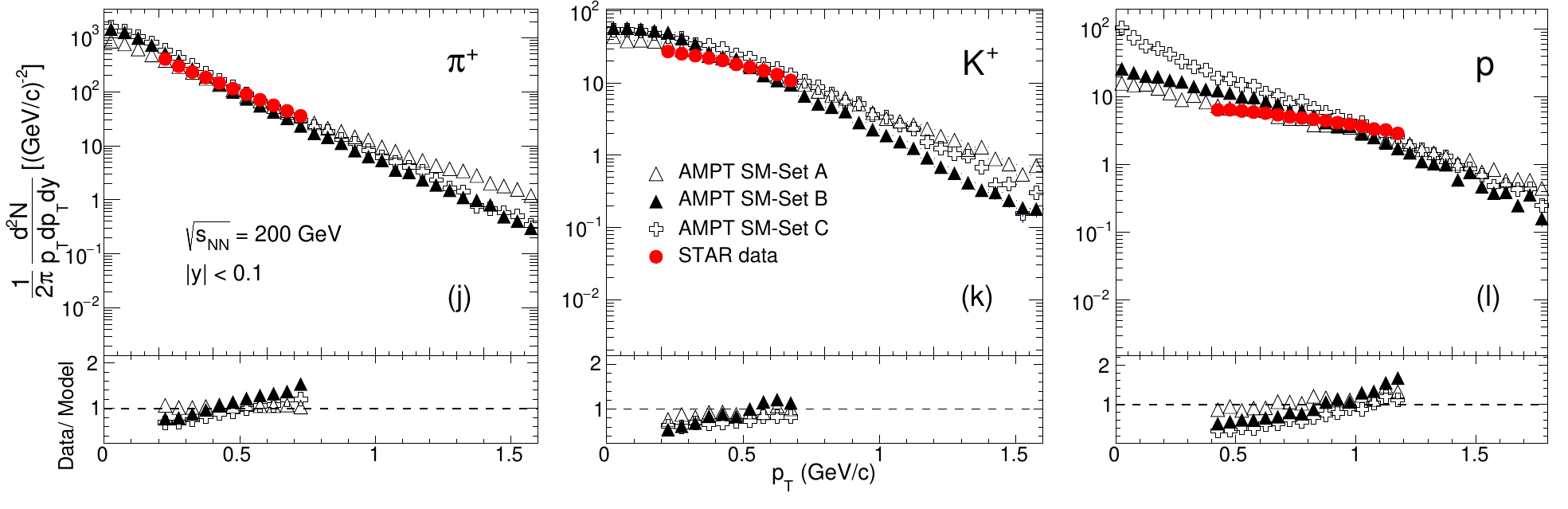}
  \caption{Midrapidity ($|y| < 0.1$) transverse momentum spectra of
    $\pi^{+}$, $K^{+}$, $p$ at $\sNN$ = 7.7 and 200 GeV for 0-5\%
    central \Au collisions from default [(a)--(f)] and string melting
    [(g)--(l)] versions of the AMPT model using parameter sets A, B,
    and C. Experimental data from the STAR
    Collaboration~\cite{Adamczyk:2017iwn, Abelev:2008ab} are shown by
    solid circles. The data-to-model ratios are presented at the bottom of each panel.}
  \label{fig:3}
\end{center}
\end{figure*}
Figure~\ref{fig:1} shows the invariant yield versus \pT in \Au
collisions at $\sNN$ = 27 GeV for positively charged particles
($\pi^{+}$, $K^{+}$, p). The results are shown using the set B
parameters for representation. 
The top three panels  [(a)--(c)] 
represent the results for the default AMPT version while the AMPT string
melting results are shown in the bottom three panels [(d)--(f)] . Results from the nine collision centralities 0-5\%, 5-10\%, 10-20\%, 20-30\%, 30-40\%, 40-50\%, 60-70\%, and 70-80\% are shown. The invariant yield decreases with increasing \pT and also while going from central to peripheral collisions. On comparing the inverse slopes of the spectra for three particles, we observe that they follow the order $p>K>\pi$. The same behavior is observed at 7.7 and 200 GeV and for all parameter sets. 
The negatively charged particles (not presented here) also show similar behavior.

Figure~\ref{fig:3} compares the \pT spectra of $\pi^+$, $K^+$, and $p$ for both versions of the AMPT model and using the three different parameter sets with experimental data in \Au collisions at $\sqrt{s_{NN}} =$ 7.7 and 200 GeV for 0--5\% collision centrality. 
The top six panels  [(a)--(f)]  represent the comparison of default
model with the experimental data, while the six panels at the bottom
[(g)--(l)]  represent the same for the string melting version. The data-to-model ratios are shown at the bottom of each plot.
For the default version, at 7.7 GeV, set B parameters describe the $\pi^{+}$ spectra better. Both the $K^+$ and $p$ spectra are described better by the set A parameters at this energy. 
At 27 GeV (plots not presented here), the $\pi^+$ spectra is described well by set C parameters. The $K^+$ and $p$ spectra are explained better by set A parameters.
At 200 GeV, the set A and B parameters describe the $\pi^{+}$ and $K^+$ while set A describes the $p$ spectra better as compared to the other sets. 

For string melting, at 7.7, 
27 (plots not presented here), and 200 GeV, set A parameters describe the $\pi^{+}$ and $p$ spectra well for 0--5\% centrality. 
The $K^{+}$ spectra at 7.7 GeV are underpredicted by all sets by about a factor of 2, with set A parameters showing a better \pT dependence.
At 
27 GeV, the data-to-model ratio comes closer to unity for set A parameters
but is still underpredicted. At 200 GeV, the ratio of data to model for $K^{+}$ becomes less than unity.
Thus, the ratio of data to model for $K^{+}$  decreases with increasing energy from about 2 at 7.7 GeV to just less than unity at 200 GeV using set A parameters. 
This suggests that the string melting version is important for description of kaons towards higher center-of-mass collision energies but does not characterize lower energy collisions well. 

{\it 
To summarize the observations from Fig.~\ref{fig:3}:
\begin{itemize}
\item
The pion spectrum at 7.7 GeV is described well by SM model set A parameters. 
At 27 GeV, it is described better by default set A paramaters.
At 200 GeV, it is described by both default and SM set A parameters.
\item
The kaon spectra at 7.7 and 
27 GeV are described better by default set A parameters. 
At 200 GeV, they are described adequately by default set A parameters but are
slightly overestimated. 
\item
The proton spectra at 7.7 
and 27 GeV are described well by SM set A parameters at low \pT and by default set A parameters at high \pT. 
At 200 GeV, the spectra are described adequately by both default and SM set A parameters.
\end{itemize}
}

The spectra comparison are quantified by comparing particle yields, average transverse momenta, and particle ratios.

 \label{Sec:Spectra}

\subsection{Particle yields (dN/dy)}\label{Sec:Yield}
\begin{figure*}[!]
\begin{center}
  \includegraphics[width = \textwidth]{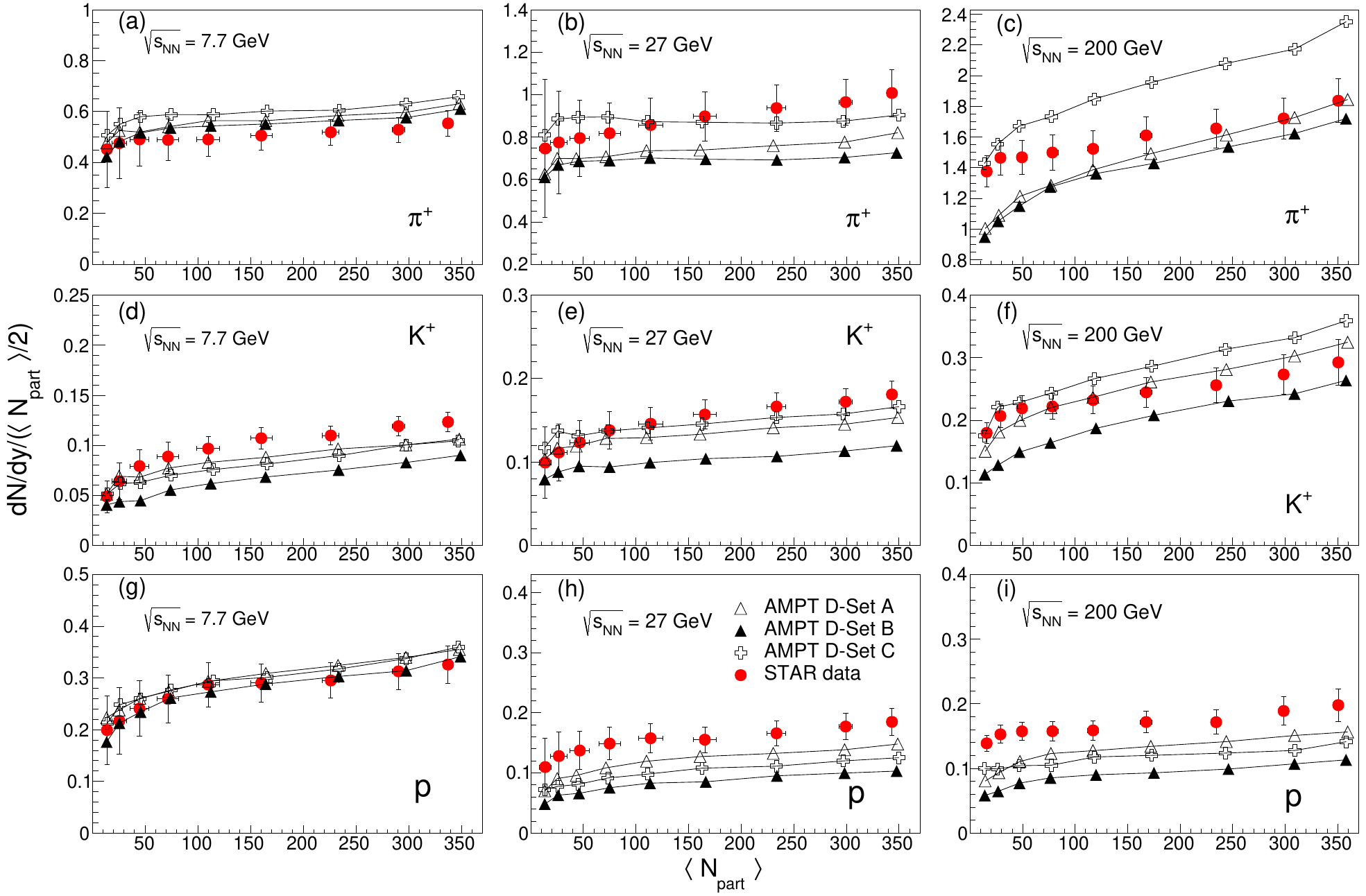}
  \includegraphics[width = \textwidth]{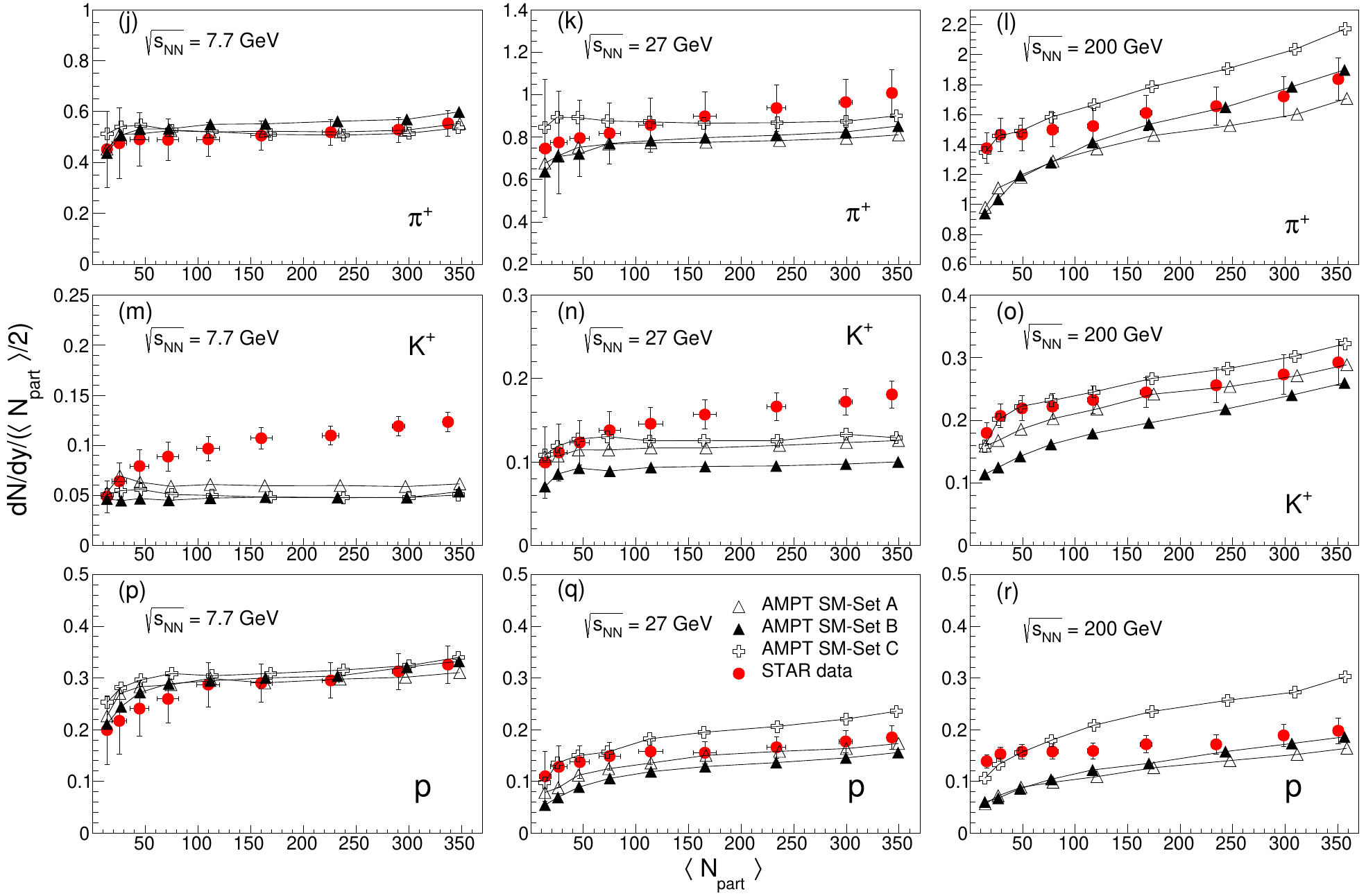}
  \caption{Centrality dependence of $dN/dy$ normalized by half participant $\langle N_{\mathrm{part}}\rangle /2$ for positive particles at mid-rapidity ($|y| < 0.1$) in \Au collisions at $\sNN$ = 7.7, 27, 200 GeV from the AMPT default [(a)--(i)] and string melting [(j)--(r)] models. Results are presented using the parameter sets A, B, and C. Experimental data from the STAR Collaboration\cite{Adamczyk:2017iwn,Abelev:2008ab} are shown by solid circles.}
  \label{fig:4}
\end{center}
\end{figure*}
\begin{figure*}[!]
\begin{center}
  \includegraphics[width = \textwidth]{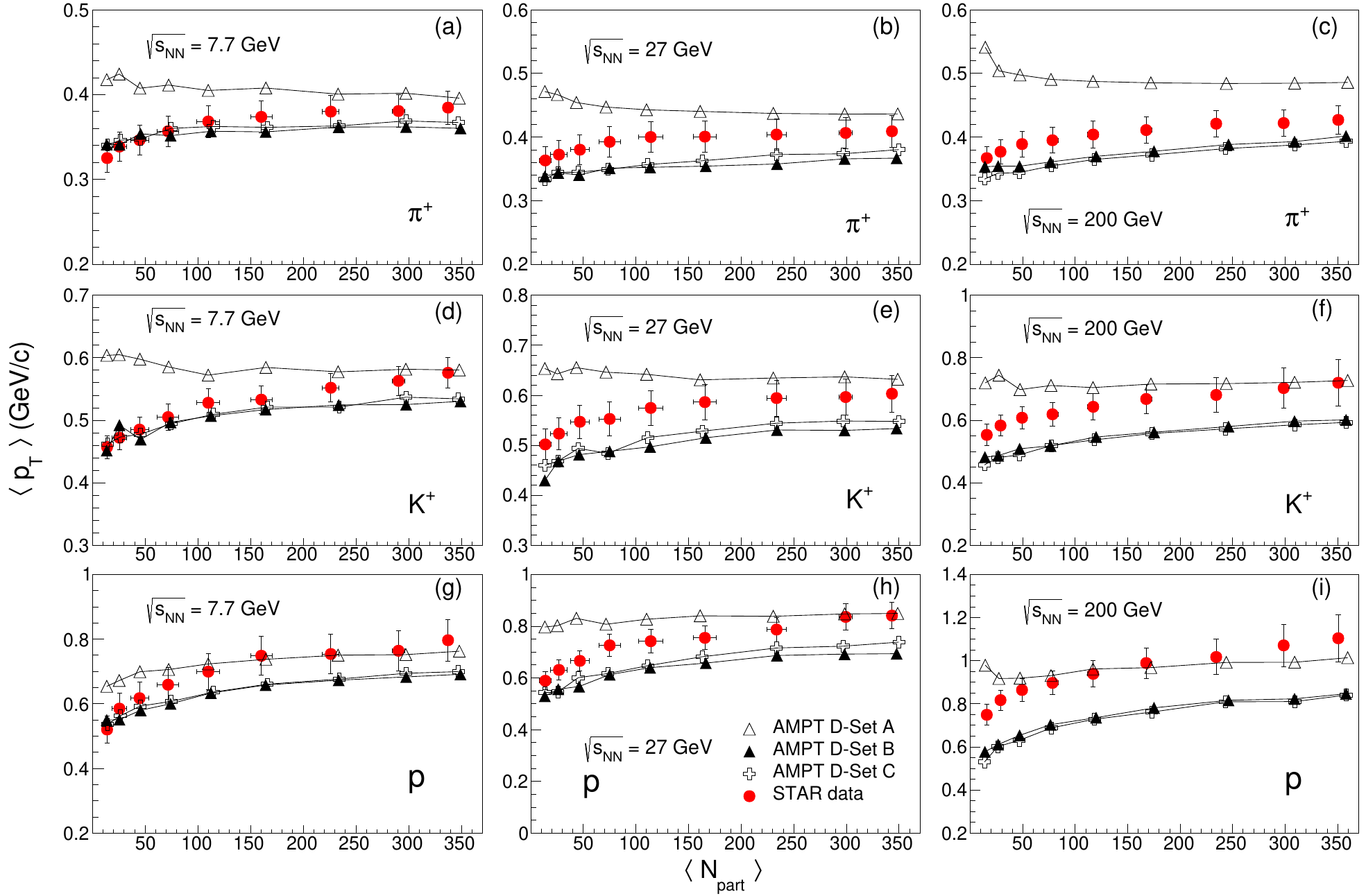}
  \includegraphics[width = \textwidth]{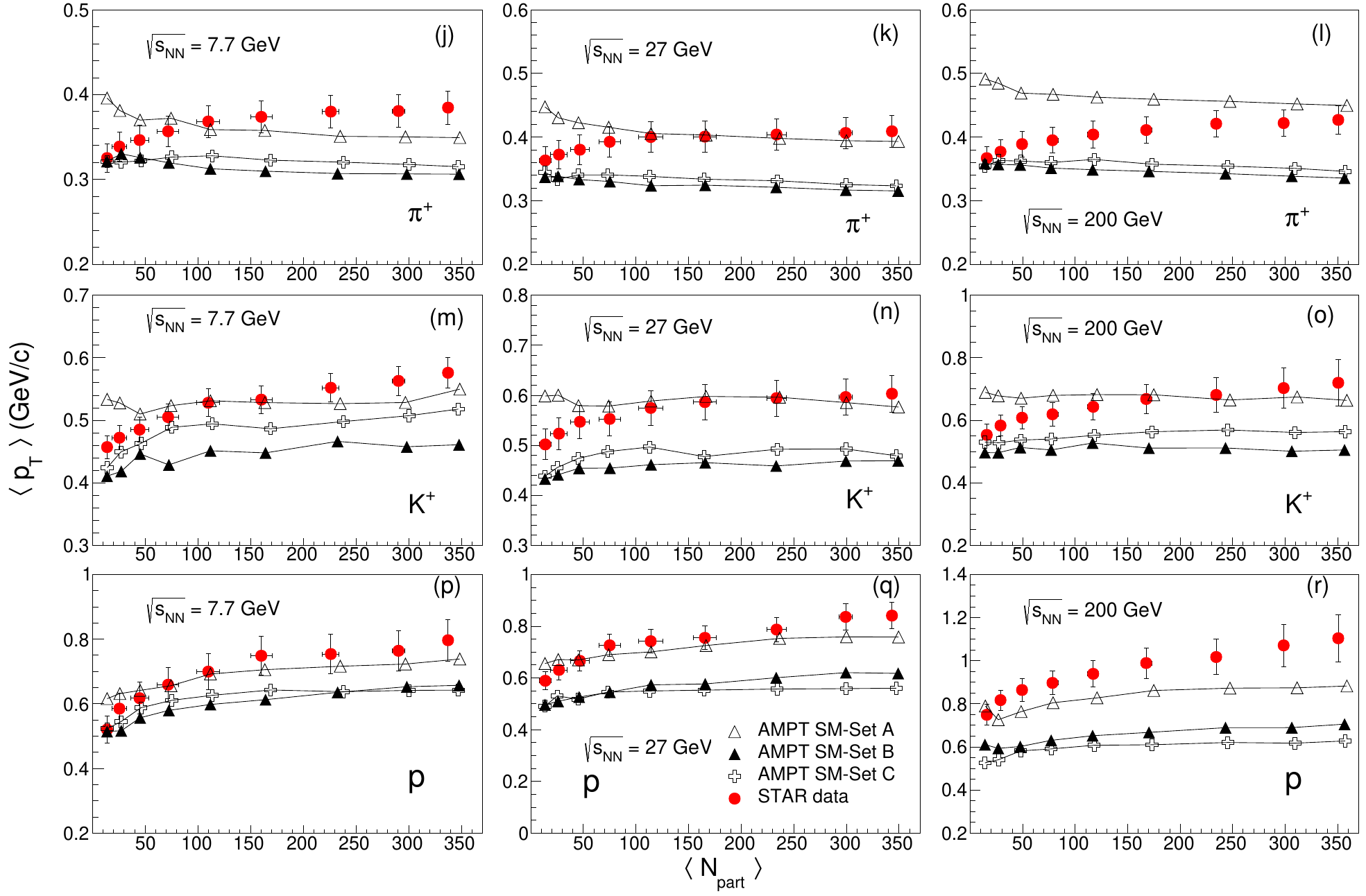}
  \caption{Centrality dependence of \pTavg for positive particles at mid-rapidity ($|y| < 0.1$) in \Au collisions at $\sNN$ = 7.7, 27, and 200 GeV from the AMPT default  [(a)--(i)]  and string melting [(j)--(r)] models. Results are presented using the parameter sets A,B, and C. Experimental data from the STAR Collaboration\cite{Adamczyk:2017iwn,Abelev:2008ab} are shown by solid circles.}
  \label{fig:5}
\end{center}
\end{figure*}
Figure~\ref{fig:4} shows the centrality dependence of yield $dN/dy$
normalized by half the number of participants, $\langle
N_{\mathrm{part}} \rangle/2$, for  $\pi^+$, $K^+$, and protons in Au+Au
collisions at 7.7, 27, and 200 GeV. The results from the default
version are shown in the top three rows  [(a)--(i)], while those using
the string melting version are shown in the bottom three rows [(j)--(r)]. The results using the three sets of parameters in both the model versions are compared with the experimental data. The experimental data show an increase of yield from peripheral to central collisions suggesting particle production by both soft and hard processes. 

In the default version, the $(dN/dy)/(0.5\langle N_{\rm{part}} \rangle)$ of $\pi^{+}$ 
at 7.7 GeV is described by set B parameters at all $\langle N_{\rm{part}} \rangle$ values. At 27 GeV, set C parameters agree with data at all $\langle N_{\rm{part}} \rangle$ values, but $\langle N_{\rm{part}} \rangle$ dependence is flat as opposed to the data in which it increases from peripheral to central collisions. At 200 GeV, none of the sets could explain the behavior observed in data for all $\langle N_{\rm{part}}\rangle $ values. The set A parameters could only describe the data for $\langle N_{\rm{part}}\rangle>$ 100 while set C parameters agree with data for $\langle N_{\rm{part}} \rangle <$ 40. 
The $K^+$ 
yields at 7.7 GeV are not explained by any of the parameter sets for all $\langle N_{\rm{part}} \rangle $. 
The set A parameters can only describe the data for $\langle N_{\rm{part}} \rangle <$ 120. 
At 27 GeV, $K^+$  yields are better described by set C parameters for all $\langle N_{\rm{part}} \rangle$, while at 200 GeV, the set A parameters describe the  $K^+$ yields for all $\langle N_{\rm{part}} \rangle $. 
The proton  
yields are described by all the parameter sets at all $\langle N_{\rm{part}} \rangle $ for 7.7 GeV, 
but none of them work for 27 GeV other than sets A and C at $\langle N_{\rm{part}} \rangle <$ 30, 
whereas at 200 GeV none of the parameters could explain the $p$ yields at any centrality.    

For the AMPT model with string melting, the $(dN/dy)/(0.5\langle N_{\rm{part}} \rangle)$ of $\pi^{+}$ 
at 7.7 GeV is described by all the parameters at all $ N_{\rm{part}}$ values. 
However, the set C parameters show a rather flat behavior as opposed to the slight increase from peripheral to central collisions. 
At 27 GeV, the set C parameters describe the $\pi^+$ yields at all $\langle N_{\rm{part}} \rangle$ values but set A and B parameters are closer in agreement with the data in peripheral collisions. 
At 200 GeV, in central collisions ($\langle N_{\rm{part}} \rangle > 100$), pion yields are well described by set B parameters while those in peripheral collisions ($\langle N_{\rm{part}} \rangle < 130$) are described by set C parameters. 
The $K^{+}$ 
yields are only described by set A parameters below $\langle N_{\rm{part}} \rangle$ 50 at 7.7 GeV, 
below $\langle N_{\rm{part}} \rangle$ 130 by set C parameters at 27 GeV, and for all $\langle N_{\rm{part}} \rangle$ by set C parameters at 200 GeV. 
The proton 
yields at 7.7 GeV are described by all parameter sets at all $\langle N_{\rm{part}} \rangle$, 
at 27 GeV by set A parameters at all $\langle N_{\rm{part}} \rangle$, 
and at 200 GeV by set B parameters for $\langle N_{\rm{part}} \rangle>$ 220 and by set C parameters for $\langle N_{\rm{part}} \rangle<$ 90 but not by any parameter set at the most peripheral point. 

{\it To summarize the observations for all centralities:
\begin{itemize}
\item
The pion yield is described 
by set C parameters for $\sqrt{s_{NN}} \leq$ 27 GeV for the SM model, but by none of the models at 200 GeV. However, the 200 GeV pion yield is constrained between sets A and C at all $\langle N_{\rm{part}} \rangle$ for both versions of AMPT.
\item
The kaon yield at 7.7 GeV is not explained at all $\langle
N_{\rm{part}} \rangle$ by any set with either versions (the models
underestimate the data), explained at 27 GeV by the default model with
set C parameters and also at 200 GeV by the default model with set A
parameters and by the SM model with set C parameters. Thus, at 7.7
GeV, the kaon (strange particle) production is not explained by the AMPT model.
\item
The proton yield at 7.7 GeV is explained by all parameter sets with both the models, at 27 GeV by set A parameters with the SM model, but by none of the models at 200 GeV. However, the 200 GeV proton yield is constrained between sets B and C at all $\langle N_{\rm{part}} \rangle$ for the AMPT SM version.
\item
In general, for most cases, it is observed that the set C parameters corresponding to 
 largest $a=2.2$ give higher yields while set B parameters corresponding to the smallest $a=0.5$ give smaller yields, as expected. 
\end{itemize}
} 

\subsection{Average transverse momentum (\pTavg)}
\label{Sec:ptaverage}
Figure~\ref{fig:5} shows the centrality dependence of average
transverse momentum \pTavg for  $\pi^+$, $K^+$, and protons in Au+Au
collisions at $\sNN =$ 7.7, 27, and 200 GeV. The results from the
default version are shown in the top three rows [(a)--(i)]  while
those from the string melting version are shown in the bottom three
rows [(j)--(r)]. The results using the three parameter sets A, B, and C are
compared with experimental data using both the default and string
melting versions. The data show increase of \pTavg from peripheral to central collisions suggesting increasing radial flow towards central collisions. The \pTavg reflects the shape (slope) of the spectra. 

Using the default version, \pTavg of $\pi^{+}$ 
at 7.7 GeV is described by set C parameters for all $\langle N_{\rm{part}} \rangle$. 
At 27 GeV, set A and set C parameters agree with data at $\langle N_{\rm{part}}\rangle>$ 220. 
While the set A parameters do not follow the behavior of data, set B and C reproduce the data qualitatively and tend to agree with them at the last two peripheral points. 
At 200 GeV, none of the sets could explain the behavior observed in data for all $\langle N_{\rm{part}}\rangle $ values. The set B parameters only describe the most peripheral data. 
The $K^+$ \pTavg at 7.7 GeV can only be explained by set A parameters for $\langle N_{\rm{part}} \rangle>$ 220, and by sets B and C for $\langle N_{\rm{part}} \rangle<$ 170. 
 At 27 GeV, $K^+$ \pTavg are better described by set A parameters for $\langle N_{\rm{part}} \rangle>$ 150.
Set A shows a flat behavior with $\langle N_{\rm{part}} \rangle$. 
 However, sets B and C only qualitatively describe the experimental data.
 At 200 GeV, set A parameters describe the $K^{+}$ 
\pTavg for $\langle N_{\rm{part}} \rangle >$ 150. Both set B and C parameters underestimate the data at all $\langle N_{\rm{part}} \rangle$. 
 For protons at 7.7 GeV, 
 \pTavg are described by set A parameters for $\langle N_{\rm{part}} \rangle >$ 50 and by both sets B and C below $\langle N_{\rm{part}} \rangle \approx$ 80. 
 At 27 GeV, set A parameters describe the protons' \pTavg for $\langle N_{\rm{part}} \rangle >$ 220. For peripheral collisions, set B and C parameters give closer \pTavg values to experimental data but underestimate nevertheless. 
 At 200 GeV, the set A parameters could explain the $p$ 
\pTavg for all $\langle N_{\rm{part}} \rangle$ values except the two peripheral bins. 
 The other two parameter sets underestimated the data quite significantly.

For AMPT string melting, the \pTavg of $\pi^{+}$ 
 at 7.7 GeV is described by set A parameters at three mid-central collisions but is underestimated (overestimated) at central (peripheral) collisions. Sets B and C can only describe the data at the last three peripheral bins.
At 27 GeV, the set A parameters could explain the data for $\langle N_{\rm{part}} \rangle \geq$ 70 while set B and C parameters could only agree with data at the most peripheral bin. 
Increasing the energy further to 200 GeV leads to the overestimation of data by set A parameters with only the most central point being sufficiently close to the data. Set C can describe the data at the three most peripheral points and set B at the two most peripheral points.  
The $K^+$ \pTavg at 7.7 GeV are described by set A parameters for four mid-central points but are underestimated (overestimated) at central (peripheral) collisions. The set C parameters tend to describe the data for $\langle N_{\rm{part}} \rangle <$ 90. 
Increasing the energy to 27 GeV, for $K^+$, leads to better agreement also in central collisions by set A parameters. These parameters describe the data for all but the last two most peripheral $\langle N_{\rm{part}} \rangle$ values. 
Increasing the energy further to 200 GeV, for $K^{+}$, 
does not change the results much for set A parameters which still describe the data from mid-central to central collisions. Using set C parameters for $K^+$, the model agrees with data at the most peripheral point. 
The proton 
\pTavg at 7.7 GeV are described by set A parameters at all $\langle N_{\rm{part}} \rangle$ except at the most peripheral bin. The set C parameters seem to describe the data at peripheral collisions below $\langle N_{\rm{part}} \rangle \approx$ 100.
At 27 GeV, the set A parameters describe the proton data at all but the two most central points and the most peripheral point. 
The other two parameter sets underestimate the data. 
At 200 GeV, the set A parameters only describe the proton data at most peripheral bin and 
underestimate the data for all other $\langle N_{\rm{part}} \rangle$ values. 
The sets B and C underestimate the data at all $\langle N_{\rm{part}} \rangle$ values. 

{\it To summarize the above observations:
\begin{itemize}
\item
The \pTavg of pions at 7.7 GeV is described at all $\langle N_{\rm{part}} \rangle$ by default AMPT set C parameters. At 27 GeV, it is described by AMPT SM set A parameters for only $\langle N_{\rm{part}} \rangle > $ 50 and is constrained between sets A and C below that. At 200 GeV, it is explained by none of the models but constrained between sets A and B for both the default and SM versions.
\item
The \pTavg of kaons at 7.7 GeV is described partially by default AMPT set A parameters for $\langle N_{\rm{part}} \rangle > $ 220, and by default AMPT set B and C parameters for $\langle N_{\rm{part}} \rangle < $ 170.
At 27 GeV, it is explained by SM set A parameters for all $\langle N_{\rm{part}} \rangle$ except at the two most peripheral points. For the two most peripheral bins, it is constrained between SM sets A and C.
At 200 GeV, it is explained by default and SM set A parameters for $\langle N_{\rm{part}} \rangle >$ 100. Below that, it is constrained better between SM sets A and C.
\item
The proton \pTavg at 7.7 GeV is described by SM set A parameters at all $\langle N_{\rm{part}} \rangle$ except the most peripheral bin. The SM sets B and C describe the peripheral bin. 
At 27 GeV, again, SM set A parameters work better for all but the most peripheral bin and two most central bins. At 200 GeV, the proton \pTavg is explained at all but last two peripheral bins by default set A parameters. The last two bins are constrained between default sets A and B.
\end{itemize}
} 

\subsection{Particle ratios}\label{Sec:Ratios}
\begin{figure*}[!]
\begin{center}
  \includegraphics[width = 17cm]{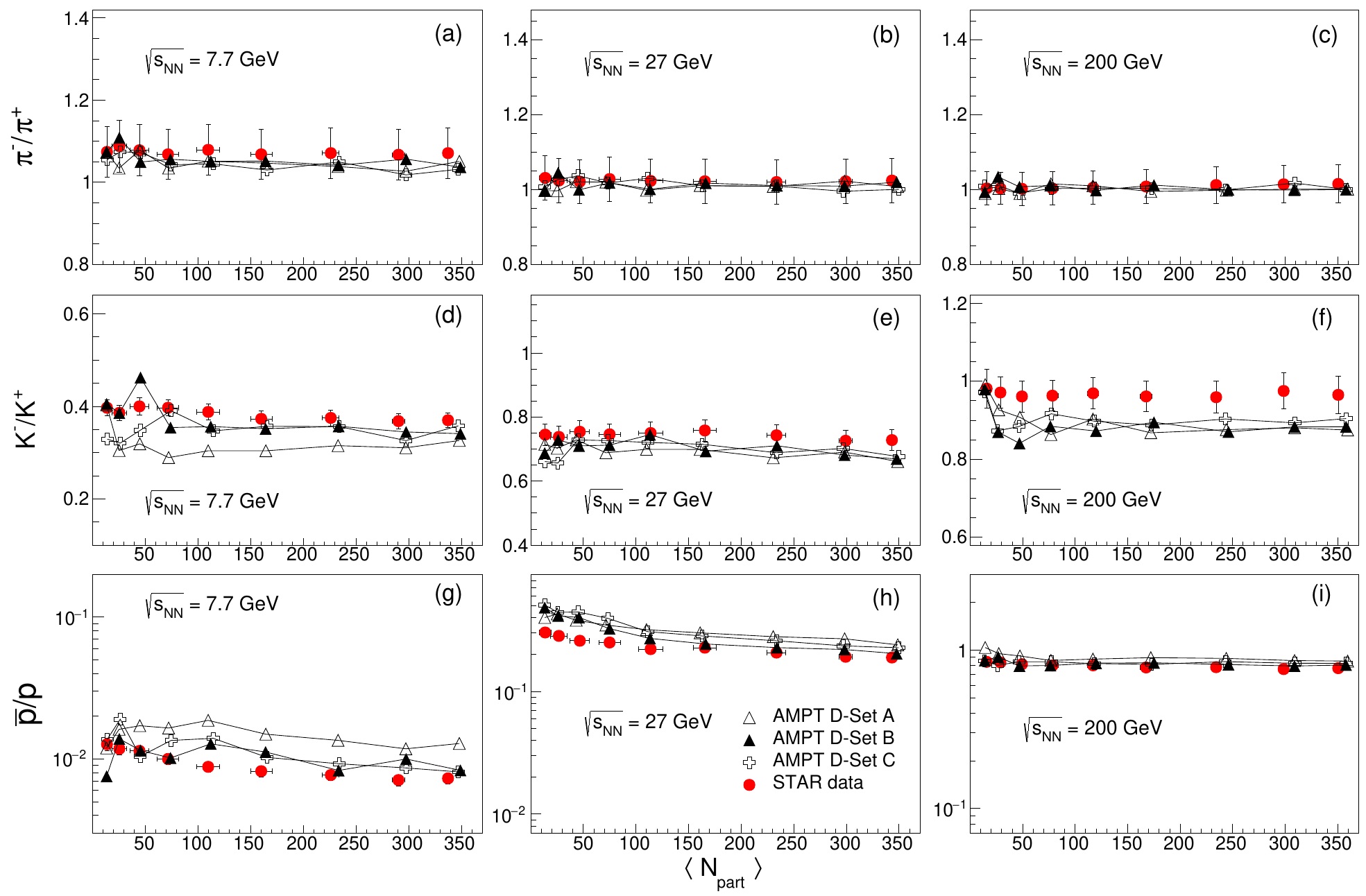}
  \includegraphics[width = 17cm]{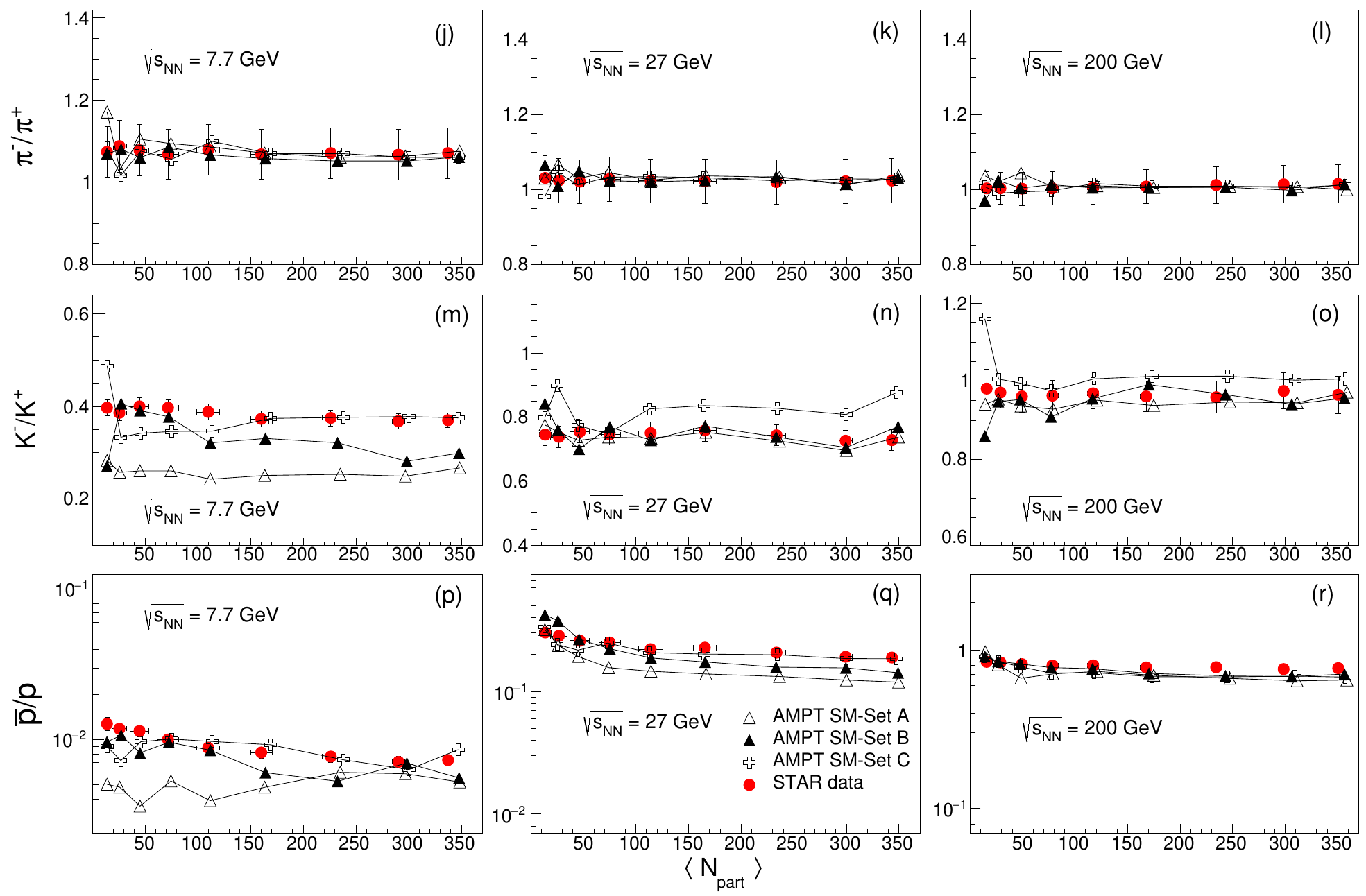}
  \caption{Centrality dependence of antiparticle-to-particle ($\pi^{-}/\pi^{+}$, $K^{-}/K^{+}$, $\bar{p}/p$) 
ratios at mid-rapidity ($|y| < 0.1$) in \Au collisions at $\sNN$ =
7.7, 27, and 200 GeV from the AMPT default [(a)--(i)]  and 
SM [(j)--(r)] models. Results are presented using the parameter sets A, B and C. Experimental data from the STAR Collaboration~\cite{Adamczyk:2017iwn,Abelev:2008ab} are shown by solid circles.}
  \label{fig:6}
\end{center}
\end{figure*}
\begin{figure*}[t]
\begin{center}
  \includegraphics[width = 17cm]{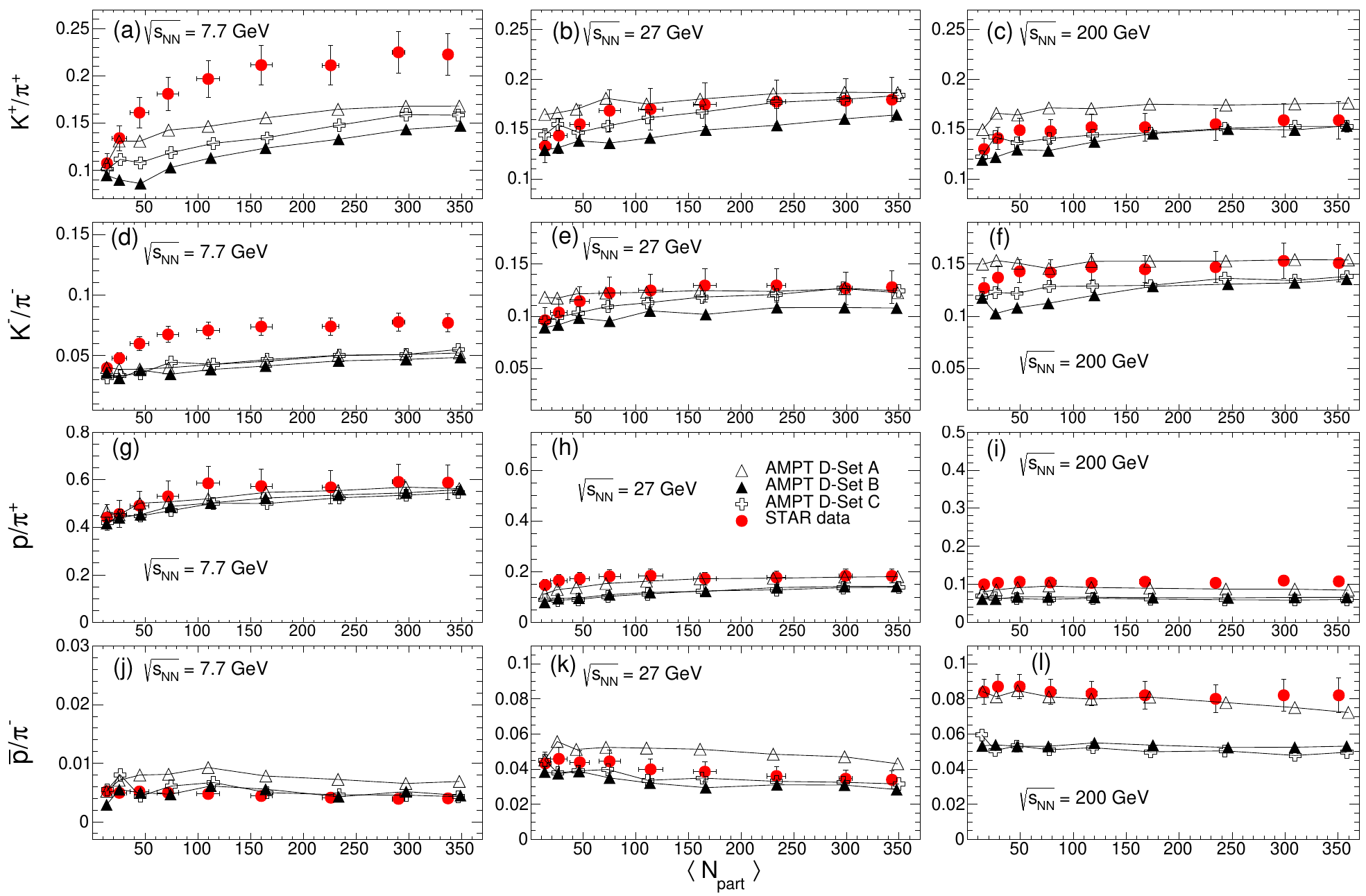}
  \includegraphics[width = 17cm]{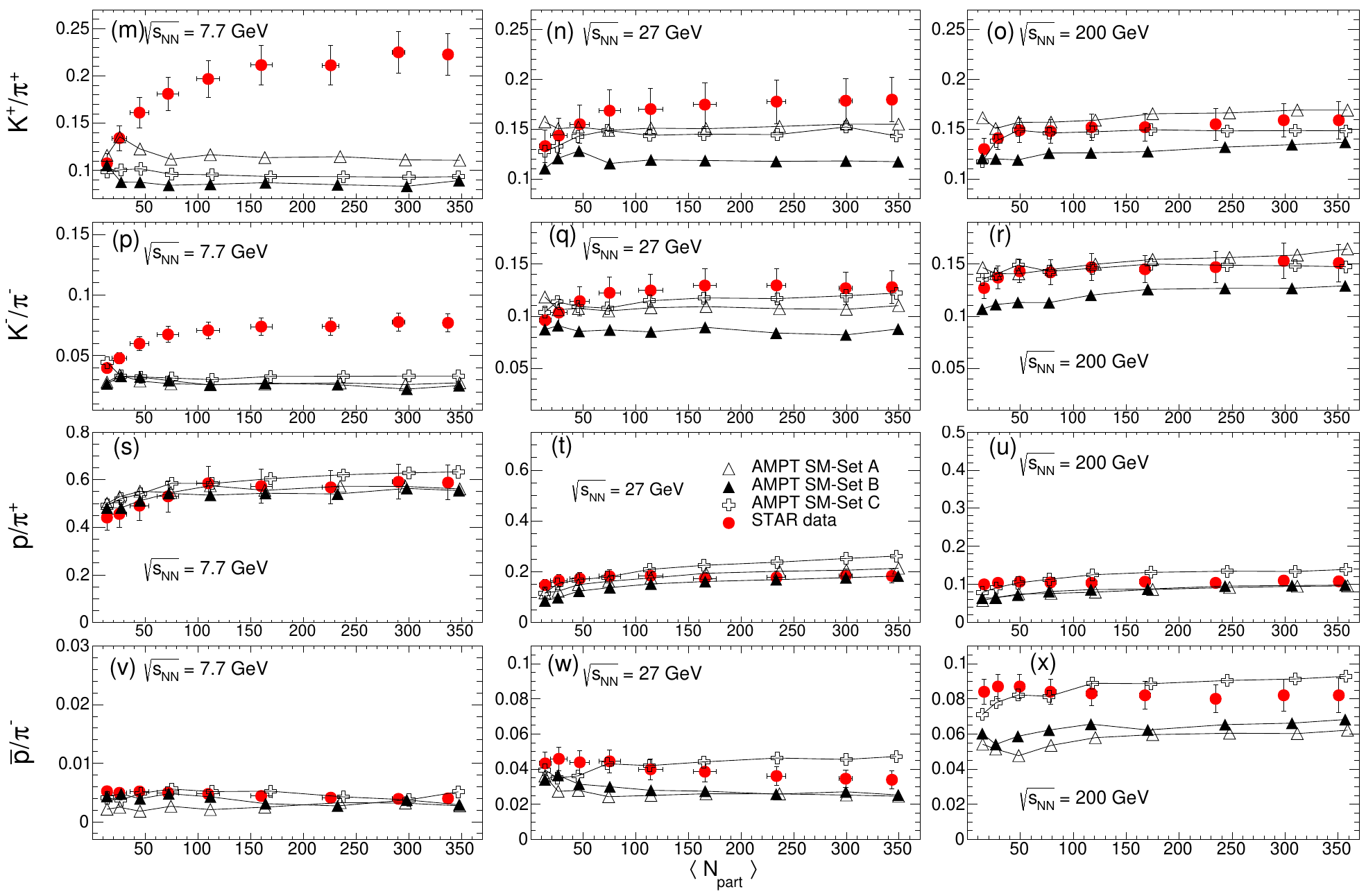}
  \caption{Centrality dependence of 
mixed particle ($K^{+}/\pi^{+}$, $K^{-}/\pi^{-}$, $p/\pi^{+}$,
$\bar{p}/\pi^{-}$) ratios at mid-rapidity ($|y| < 0.1$) in \Au
collisions at $\sNN$ = 7.7, 27, and 200 GeV from the default
[(a)--(l)] and AMPT SM [(m)--(x)] models. Results are presented using the parameter sets A,B, and C. Experimental data from the STAR Collaboration~\cite{Adamczyk:2017iwn,Abelev:2008ab} are shown by solid circles.}
  \label{fig:7}
\end{center}
\end{figure*}
 In Fig. \ref{fig:6}, we show the centrality dependence of various antiparticle-to-particle ($\pi^{-}/\pi^{+}$, $K^{-}/K^{+}$, $\bar{p}/p$) ratios 
at mid-rapidity ($|y| < 0.1$) in \Au collisions at $\sNN$ = 7.7, 27, and 200 GeV obtained from the default [(a)--(i)] and SM [(j)--(r)] AMPT models using the three parameter sets A, B, and C. The results are again compared with the corresponding experimental data.

The default AMPT model could reasonably predict the $\pi^{-}/\pi^{+}$ ratio at the three energies with all the parameter cases. 
The $K^{-}/K^{+}$ ratio at 7.7 GeV is mostly under-estimated by set A parameters while the set B and set C parameters give closer values to data in general. 
At 27 GeV, the results with the three parameter sets are close to each other and the data, agreeing marginally with the data. 
At 200 GeV, the $K^{-}/K^{+}$ ratio is mostly underestimated by three parameter sets but matches with the data in peripheral collisions.  
The $\bar{p}/p$ ratio at 7.7 GeV is mostly overestimated by all the three parameter sets. For $\langle N_{\rm{part}}\rangle<$ 90 (except the most peripheral bin), set B parameters explain the data. 
At 27 GeV, the $\bar{p}/p$ ratio is explained by set B parameters for $\langle N_{\rm{part}}\rangle>$ 100. 
At 200 GeV, all the three parameter sets seem to describe the $\bar{p}/p$ ratio, with the exception of the most peripheral point by set A parameters.  

Similar to the default model, the AMPT model with string melting could reasonably predict the $\pi^{-}/\pi^{+}$ ratio at the three energies with all the three parameter cases. 
The $K^{-}/K^{+}$ ratio at 7.7 GeV is generally described by set C parameters for central collisions $\langle N_{\rm{part}}\rangle>$ 150. Set B parameters could only explain the ratio at three points before the most peripheral bin.  
At 27 and 200 GeV, set A parameters describe the data at all centralities. The set B parameters could also explain the data at all but two centralities. 
The $\bar{p}/p$ ratio at 7.7 GeV
is described by the set C parameters for all centralities except at the two most peripheral bins. 
At 27 GeV, it is described by the set C parameters at all centralities. At 200 GeV, all three sets give similar values, close to the experimental $\bar{p}/p$ ratio.

The mixed particle ratio results could help in better differentiating among the three parameter sets. 
 In Fig. \ref{fig:7}, we show the centrality dependence of various 
mixed ($K^{+}/\pi^{+}$, $K^{-}/\pi^{-}$, $p/\pi^{+}$, $\bar{p}/\pi^{-}$) particle ratios  
at mid-rapidity ($|y| < 0.1$) in \Au collisions at $\sNN$ = 7.7, 27, and 200 GeV obtained from the default [(a)--(l)] and SM [(m)--(x)] AMPT models using the three parameter sets A, B, and C. The results are compared with the corresponding experimental data.

For default AMPT model, the $K^+/\pi^+$ ratio at 7.7 GeV is not explained by any of the parameter sets except at very peripheral collisions. 
At 27 GeV, the $K^+/\pi^+$ ratio is described by set C parameters at all $\langle N_{\rm{part}}\rangle$. 
The set A parameters describe the data at all centralities except at the most peripheral one,  
while set B parameters describe the ratio at almost all $\langle N_{\rm{part}}\rangle$ values except in mid-central collisions.
Similar conclusions could be drawn for 200 GeV except that the set A parameters now miss the data at more $\langle N_{\rm{part}}\rangle$ values.
The same as the $K^+/\pi^+$ ratio, the $K^-/\pi^-$ ratio at 7.7 GeV is also not described by any of the three parameter sets except at the very peripheral points. At 27 GeV, the ratio is well explained by set C parameters for all $\langle N_{\rm{part}}\rangle$. The set A parameters also describe the data at all $\langle N_{\rm{part}}\rangle$ except at the most peripheral bin, 
while set C parameters work well at peripheral collisions. Similar conclusions could be drawn at 200 GeV except that the set C parameters also miss a few points towards the peripheral collisions. Thus, in this case, set A describes the data better at all  $\langle N_{\rm{part}}\rangle$ except the peripheral point.
The $p/\pi^+$ ratio at 7.7 GeV is described by all parameter sets at all $\langle N_{\rm{part}}\rangle$. 
At 27 GeV, the ratio is described by set A parameters at all  $\langle N_{\rm{part}}\rangle$. 
At 200 GeV, the $p/\pi^+$ ratio predicted by set A parameters is closer to data but does not agree exactly with it. 
The $\bar{p}/\pi^-$ ratio at 7.7 GeV
is described by set B and C parameters at all $\langle N_{\rm{part}}\rangle$ except at one bin towards peripheral collisions. 
At 27 GeV, it is described well by set C parameters at all $\langle N_{\rm{part}}\rangle$ values. Set B also describes this ratio at almost all the centralities. 
At 200 GeV, the ratio is explained by set A parameters for all $\langle N_{\rm{part}}\rangle$.

For the AMPT SM model, the $K^+/\pi^+$ ratio at 7.7 GeV is not explained by any parameter set except at the most peripheral collision. It is interesting to note that no set shows even the qualitative behavior of centrality dependence observed in experimental data.
At 27 GeV, the $K^+/\pi^+$ ratio is marginally described by set A parameters for most centralities except the peripheral ones. However, the $\langle N_{\rm{part}}\rangle$ dependence is well predicted by set C parameters though they consistently underestimate the data. 
At 200 GeV, the set C parameters describe the data at all centralities. The set A parameters also describe the $K^+/\pi^+$ ratio for all centralities except at the most peripheral collisions. 
The $K^-/\pi^-$ ratio at 7.7 GeV is also not described by any of the three parameter sets except at the most peripheral point by set C. 
At 27 GeV, the ratio is well explained by set C parameters for all $\langle N_{\rm{part}}\rangle$. 
The set A parameters also result in closer values to the data at most centralities. 
At 200 GeV, set C parameters describe the data at all centralities. Set A also describes the data at all centralities except at the most peripheral bin. 
The $p/\pi^+$ ratio at 7.7 GeV is described by all parameter sets at all $\langle N_{\rm{part}}\rangle$. 
At 27 GeV, the ratio is described by set A parameters at all $\langle N_{\rm{part}}\rangle$. The set B parameters describe the data for central collisions but fail at peripheral collisions while the set C parameters describe the data at peripheral collisions failing at central collisions.
At 200 GeV, the $p/\pi^+$ ratio is described by set A and B parameters towards the central collisions ($\langle N_{\rm{part}}\rangle>$ 200) and by set C parameters towards peripheral collisions ($\langle N_{\rm{part}}\rangle<$ 150). 
The $\bar{p}/\pi^-$ ratio at 7.7 GeV is described by both set B and C parameters at almost all $\langle N_{\rm{part}}\rangle$. 
At 27 GeV, the ratio is described by set C parameters from mid-central ($\langle N_{\rm{part}}\rangle<$ 200) to peripheral collisions. 
At 200 GeV, the ratio is described by set C parameters for most $\langle N_{\rm{part}}\rangle$ except at a few centrality bins. 

{\it To summarize the observations from the two models (Figs.~\ref{fig:6} and \ref{fig:7}) :
\begin{itemize}
\item
The $\pi^{-}/\pi^{+}$ ratio is described by both default and SM models using the sets A, B and C at the three energies $\sqrt{s_{NN}}=$ 7.7, 27, and 200 GeV.
\item
The $K^{-}/K^{+}$ ratio at 7.7 GeV is better described by SM set C parameters for $\langle N_{\rm{part}}\rangle> 150$. At 27 and 200 GeV, it is described at all $\langle N_{\rm{part}} \rangle$ by SM set A parameters. 
\item
The $\bar{p}/p$ ratio at 7.7 GeV is described better by SM set C parameters for all centralities except at the last two peripheral bins. 
At 27 GeV, the ratio is described well by SM set C parameters and at 200 GeV by default set B parameters at all centralities. 
\item
The $K^{+}/\pi^{+}$ ratio at 7.7 GeV is not described well by any of the models at all centralities, except the peripheral bins. The default model gives similar centrality dependence but underpredicts the data.
At 27 GeV, this ratio is described better by default set C parameters at all $\langle N_{\rm{part}}\rangle$. 
At 200 GeV, it is explained by both default and SM set C parameters at all centralities.
Thus, at 7.7 GeV, the strange particle production is not well explained by the AMPT model.
\item
The $K^{-}/\pi^{-}$ ratio results at 7.7 GeV are similar to those of the $K^{+}/\pi^{+}$ ratio. The ratio is also not explained by any model at all centralities except at the peripheral bins. 
At 27 GeV, this ratio is described by both default and SM set C parameters. 
At 200 GeV, it is explained by SM set C parameters. 
\item
The $p/\pi^{+}$ ratio at 7.7 GeV is explained by both default and SM models with all parameter sets.
At 27 GeV, the ratio is described by 
both default and SM set A parameters at all centralities. 
However, at 200 GeV, it is not explained by a single parameter set in either model at all the centralities. For central collisions, SM set A and B parameters describe the data while for peripheral collisions SM set C parameters work better. 
\item
The $\bar{p}/\pi^{-}$ ratio at 7.7 GeV is described at most 
$\langle N_{\rm{part}}\rangle$ 
by both default and SM set B and C parameters.
At 27 GeV, it is described by default set C parameters and is well explained at 200 GeV by default set A parameters at all $\langle N_{\rm{part}}\rangle$.
\end{itemize}
}

\begin{figure*}[!]
 \includegraphics[width = 12cm]{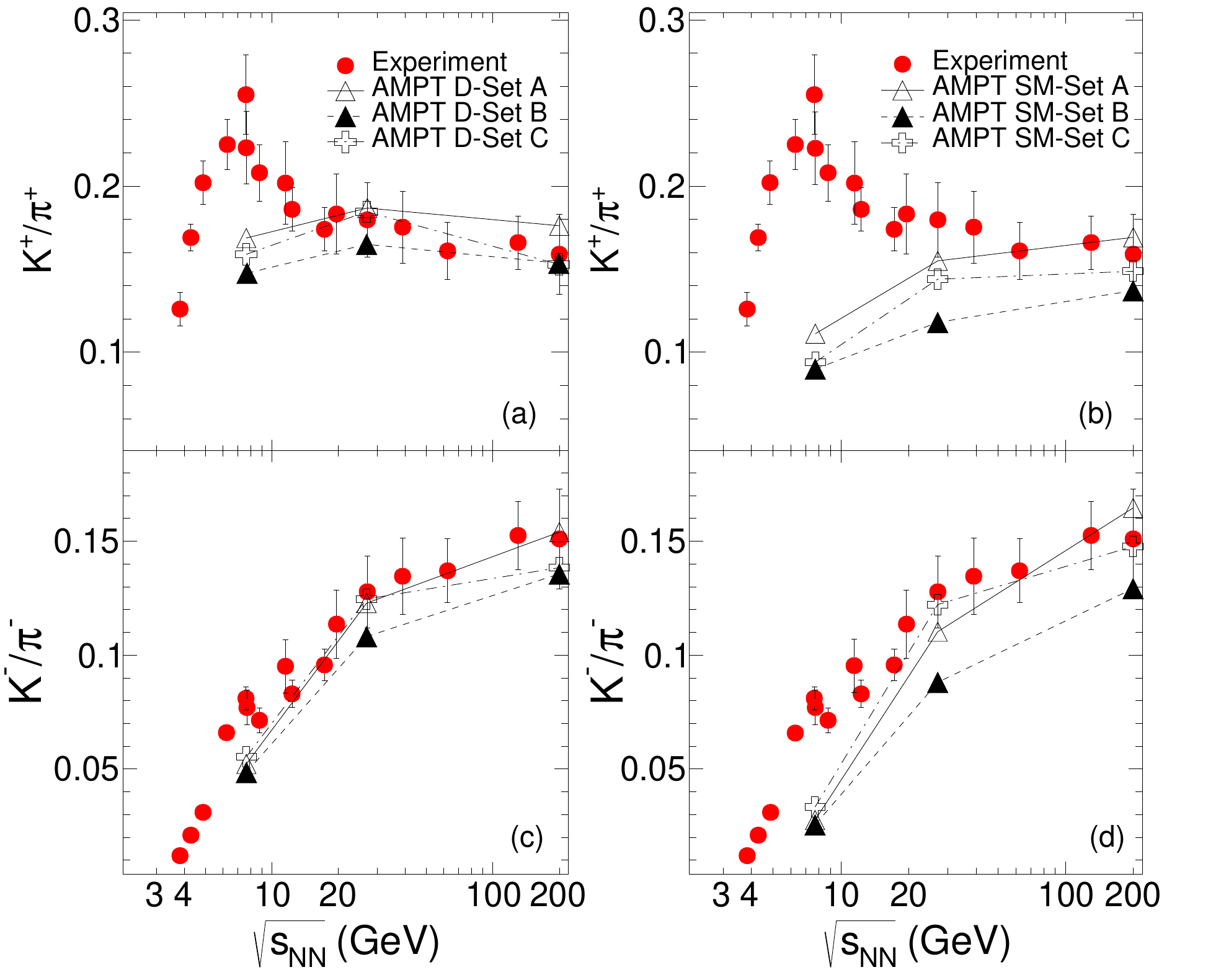}
  \caption{Comparison of $K^{\pm}/\pi^{\pm}$ ratios at mid-rapidity ($|y| < 0.1$) for 0--5\% centrality in \Au collisions at $\sNN$ = 7.7, 27,  and 200 GeV from the AMPT default [(a),(c)] and SM [(b),(d)] models with experimental data~\cite{Adamczyk:2017iwn,Abelev:2008ab,Akiba:1996xf,Ahle:1998jc,Ahle:1999uy,Ahle:1999va,Ahle:2000wq,Afanasiev:2002mx,Alt:2007aa,Abelev:2009bw}. The results from the AMPT model with parameter sets A, B, and C are presented. 
}
  \label{fig:8}
\end{figure*}
\begin{figure*}[!]
\begin{center}
  \includegraphics[width = 12cm]{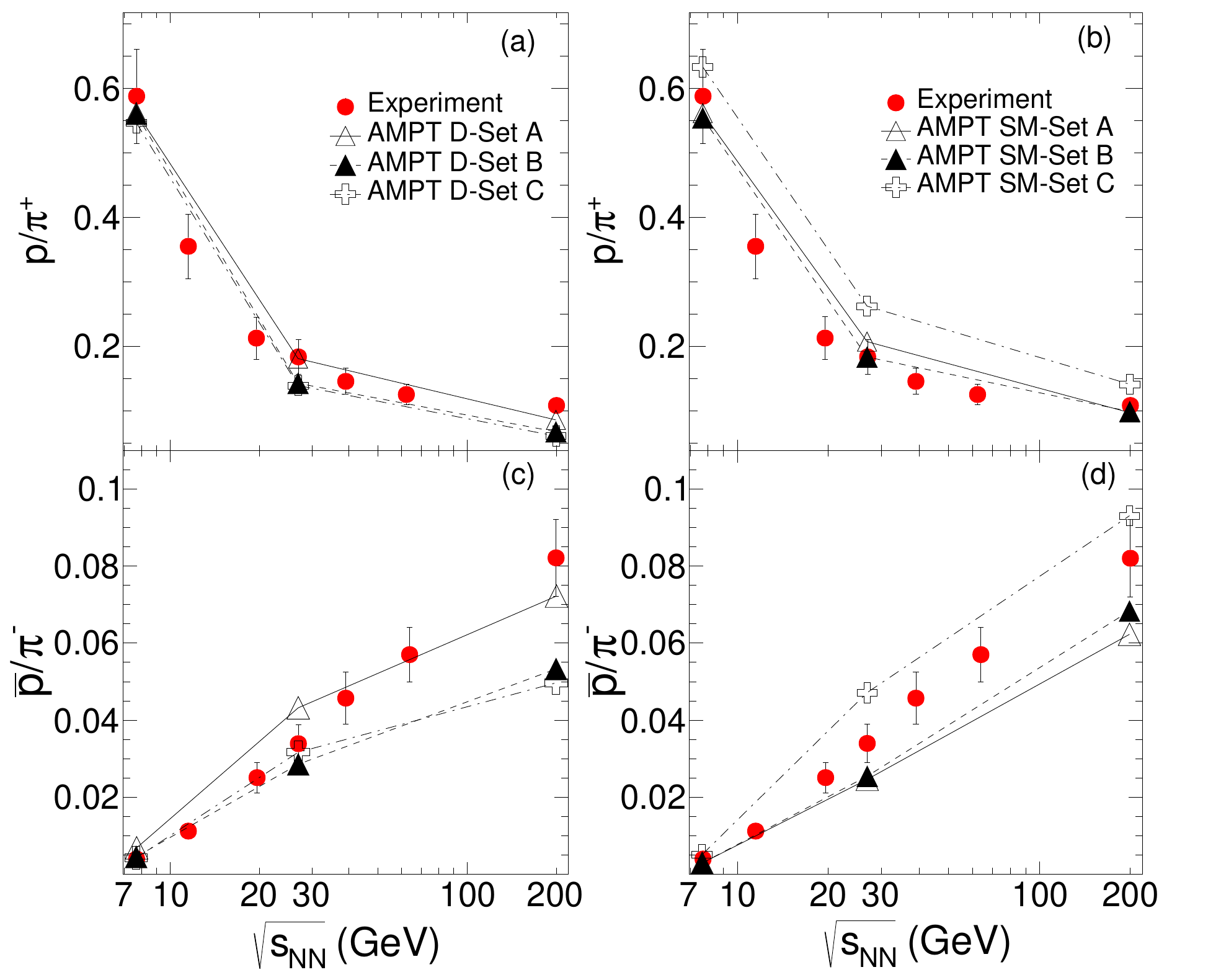}
  \caption{Comparison of $p/\pi^+$ and $\bar{p}/\pi^-$  ratios at mid-rapidity ($|y| < 0.1$) for 0--5\% centrality in \Au collisions at $\sNN$ = 7.7, 27,  and 200 GeV from the AMPT default [(a),(c)] and SM [(b),(d)]  models with experimental data~\cite{Adamczyk:2017iwn,Abelev:2008ab}. The results from the AMPT model with parameter sets A, B, and C are presented. }
  \label{fig:9}
\end{center}
\end{figure*}

\subsection{Energy dependence of particle ratios}\label{Sec:EdepRat}

 The particle yields and ratios are used in statistical thermal models to determine the freeze-out conditions in heavy-ion collisions~\cite{Cleymans:1999st,Becattini:2005xt,Andronic:2005yp,Adamczyk:2017iwn}. 
We present the energy dependence of mixed particle ratios for 0--5\% central collisions that play an important role in determining the freeze-out conditions. 

Figure ~\ref{fig:8} presents the comparison of $K^{\pm}/\pi^{\pm}$ ratios at mid-rapidity ($|y| < 0.1$) for 0--5\% centrality in \Au collisions at $\sNN$ = 7.7, 27, and 200 GeV from the AMPT default [(a),(c)] and SM [(b),(d)]  models with experimental data~\cite{Adamczyk:2017iwn,Abelev:2008ab,Akiba:1996xf,Ahle:1998jc,Ahle:1999uy,Ahle:1999va,Ahle:2000wq,Afanasiev:2002mx,Alt:2007aa,Abelev:2009bw}. The results from AMPT are presented with the parameters sets A, B, and C.
The experimental results of the $K^{+}/\pi^{+}$ ratio show an interesting trend. The ratio increases with energy, reaches a maximum, and then decreases and becomes almost constant at higher energies. It has been suggested that the peak position, also called the ``horn,'' in this energy dependence could be a signature of phase transition from hadronic to QGP gas~\cite{Alt:2007aa,Adamczyk:2017iwn}. However, the peak position also corresponds to the energy region with maximum baryon density~\cite{Randrup:2006nr}. 
For the default AMPT model, the three sets are consistent with data at 27 and 200 GeV. At 7.7 GeV, all the three sets under-predict the ratio significantly. However, among the three sets, the set A parameters are closest to the data. For SM, set A seems to be in better agreement with the data at 27 and 200 GeV but underpredicts the data at 7.7 GeV. Comparing between default and SM, the default set A parameters describe the energy dependence of the $K^{+}/\pi^{+}$ ratio better. The $K^{-}/\pi^{-}$ ratio 
at 200 GeV is described by all the three sets of the default and SM models. 
At 27 GeV, the set A and C parameters are consistent with the data. At 7.7 GeV, the ratio is again underpredicted by both versions. The default model is in closer agreement with data at lower energies. Thus, it can be concluded that strangeness (kaon) production at $\sNN$ = 7.7 GeV is not explained by the AMPT model.

Figure~\ref{fig:9} shows the comparison of $p/\pi^+$ and $\bar{p}/\pi^-$  ratios at mid-rapidity ($|y| < 0.1$) for 0--5\% centrality in \Au collisions at $\sNN$ = 7.7, 27, and 200 GeV from the AMPT default [(a),(c)] and SM [(b),(d)] models with experimental data~\cite{Adamczyk:2017iwn,Abelev:2008ab}. The results for AMPT are presented for the parameters sets A, B, and C.
In the default model, the set A parameters seem to describe the  $p/\pi^+$ ratio better at the three energies. With the SM model, both sets A and B describe the data at the three energies. 
The default AMPT set A parameters describe the $\bar{p}/\pi^-$ ratio at 7.7 and 200 GeV, while set B and C parameters describe it at 7.7 and 27 GeV. Overall, the set A parameters are closest to the data. For the SM model, the set C parameters describe the ratio at 7.7 and 200 GeV, while sets B and C only describe the data at 7.7 GeV. Again, we observe that the default AMPT model with set A parameters works better than SM model. 

In general, considering the energy dependence behavior in 0--5\% central \Au collisions,
we observe that for all observables including yields, \pTavg and
ratios, the AMPT default model with set A parameters explain the data
better than the other sets and also better than AMPT SM with all the
sets. 
However, in recent studies it has been shown that the effect of  
finite nuclear thickness of incoming colliding nuclei is important at lower energies~\cite{Lin:2017lcj,Shen:2017ruz}. So, incorporating
the finite nuclear thickness of the incoming colliding nuclei into the AMPT SM model might lead to
improvement in the results at lower energies~\cite{Lin:2017lcj}.

\section{Summary And Discussion}\label{Sec:Conclusions}
\begin{table*}[htb]
\centering
	\caption{Summary table for the performance of various sets in
          AMPT default and SM models for different observables
          presented. The $\checkmark$ symbol represents that the
          particular set describes the data well ( ``-'' symbol
          represents that the given set does not explain the data
          well at all energies as compared to other sets),
          $\ast$ represents
          that the set describes the data partially,
$X$ (in ``Remark'' column)
          suggests that none of
          the sets could describe the data. 
The figures in the parentheses followed by the symbols represent the energy $\sqrt{s_{NN}}$ (GeV) where these sets are applicable.}
        \label{tab:sum}\vspace{0.1in}
\begin{tabular}{ |c|c|c|c|c|c|c|c|c| } 

 \hline
\multicolumn{2}{|c|}{Observable} & \multicolumn{3}{|c|}{AMPT Default} & \multicolumn{3}{|c|}{AMPT  String  Melting}  & \multicolumn{1}{|c|}{Remark}\\
\cline{3-8}
\multicolumn{2}{|c|}{} & Set A & Set B & Set C & Set A & Set B & Set C&\\%
\hline

\hline
\multirow{3}{2 cm}{$p_T$ spectra (0--5\%)} & $\pi^+$ &$\checkmark$ (27, 200) & - & - & $\checkmark$ (7.7,  200)  & - & - &  \\%
\cline{2-9}

& $K^+$ & $\checkmark$ (7.7, 27)  & - & - & -& - & -&  \\%
& & $\ast$ (200)  &  &   &  &  &  & \\%
\cline{2-9}

& $p$ & $\checkmark$ (7.7, 27, 200) & - & - & $\checkmark$ (7.7, 27, 200)& - & - &  \\%
\hline

\hline
\multirow{3}{2 cm}{$dN/dy$ vs. $\langle N_{\rm{part}} \rangle$} & $\pi^+$ & $\ast$ (200) & - & - & - & $\ast$ (200) &  $\checkmark$ (7.7, 27) & \\%
\cline{2-9}

& $K^+$ & $\checkmark$ (200) & - & $\checkmark$ (27) & - & - & $\checkmark$ (200) & $X$ (7.7) \\%
\cline{2-9}

& $p$ & $\checkmark$ (7.7) & $\checkmark$ (7.7) & $\checkmark$ (7.7) & $\checkmark$ (7.7, 27)& $\checkmark$ (7.7) & $\checkmark$ (7.7) &  \\%
&  &  & & & & $\ast$ (200) &  &   \\%
\hline

\hline
\multirow{3}{2 cm}{$\langle p_T \rangle$ vs. $\langle N_{\rm{part}} \rangle$} 

& $\pi^+$ & - & - &  $\checkmark$ (7.7) &  $\ast$ (27)& - & - & $X$ (200) \\%
\cline{2-9}

& $K^+$ & $\ast$ (7.7, 200) & $\ast$ (7.7)  & $\ast$ (7.7)  & $\ast$ (27, 200)  & - & - &\\%
\cline{2-9}

& $p$ & $\ast$ (200) & - & - & $\ast$ (7.7, 27) & $\ast$ (7.7) & $\ast$ (7.7) &\\%
\hline

\hline
\multirow{7}{2 cm}{Ratios vs. $\langle N_{\rm{part}} \rangle$} 

& $\pi^-/\pi^+$ & $\checkmark$ (7.7, 27) & $\checkmark$ (7.7, 27) & $\checkmark$ (7.7, 27) & $\checkmark$ (7.7, 27) & $\checkmark$ (7.7, 27) & $\checkmark$ (7.7, 27) &\\%
&  & $\checkmark$ (200) & $\checkmark$ (200) & $\checkmark$ (200) & $\checkmark$ (200) & $\checkmark$ (200) & $\checkmark$ (200) &\\%
\cline{2-9}

& $K^-/K^+$ & - & - & - & $\checkmark$ (27, 200) & - & $\ast$ (7.7) &\\%
\cline{2-9}

& $\bar{p}/p$ & - & $\checkmark$ (200) & - & - & - & $\ast$ (7.7)  &\\%
& & &  & & & & $\checkmark$ (27)  &\\%
\cline{2-9}

& $K^+/\pi^+$ & - & - & $\checkmark$ (27, 200) & - & - & $\checkmark$ (200)  & $X$(7.7) \\%
\cline{2-9}

& $K^-/\pi^-$ & - & - & $\checkmark$ (27) & - & - & $\checkmark$ (27, 200) &$X$(7.7) \\%
\cline{2-9}

& $p/\pi^+$ & $\checkmark$ (7.7, 27) & $\checkmark$ (7.7) & $\checkmark$ (7.7) & $\checkmark$ (7.7, 27) & $\checkmark$ (7.7) & $\checkmark$ (7.7) &\\%
&  &  &  & &  $\ast$ (200)  &  $\ast$ (200)  &  $\ast$ (200)  &\\%
\cline{2-9}

& $\bar{p}/\pi^-$ & $\checkmark$ (200) & $\ast$ (7.7)  & $\ast$ (7.7) & - & $\ast$ (7.7) & $\ast$ (7.7) &\\%
&  & &   & $\checkmark$ (27) & &  &  &\\%
\hline

\hline
\multirow{4}{2 cm}{Ratios vs. $\sqrt{s_{NN}}$ (0-5\%) } 

& $K^+/\pi^+$  &  $\ast$ (27, 200) & $\ast$ (27, 200)  & $\ast$ (27, 200) & $\ast$ (27, 200) &  $\ast$ (200) & $\ast$ (200) & At 7.7 GeV\\%
\cline{2-8}

& $K^-/\pi^-$ & $\ast$ (27, 200) & $\ast$ (27, 200) & $\ast$ (27, 200) & $\ast$ (27, 200) &  $\ast$ (200) & $\ast$ (27, 200) & default\\%
\cline{2-8}

& $p/\pi^+$ & $\checkmark$ (7.7, 27)  & $\ast$ (7.7, 27) & $\ast$ (7.7, 27) & $\checkmark$ (7.7, 27)  & $\checkmark$ (7.7, 27) & $\ast$ (7.7) & model is\\%
&  & $\checkmark$ (200)  &  &  & $\checkmark$ (200)  & $\checkmark$ (200)  & & closer to\\%
\cline{2-8}

& $\bar{p}/\pi^-$ &  $\ast$ (7.7, 200) & $\ast$ (7.7, 27) & $\ast$ (7.7, 27) &  $\ast$ (7.7) & $\ast$ (7.7) &  $\ast$ (7.7, 200)& expt. data\\%
\cline{2-9}

\hline

\end{tabular}
\end{table*}

This study is an attempt to make the first detailed comparison of light
hadron production from the AMPT model with experimental data for three
different energies at RHIC, at different centralities and for various identified particles.
The default and SM AMPT models were initiated with different sets of
parameters (as given in Table~\ref{tab:lsff}) and the results obtained
were compared with the data from the STAR experiment. For this study,
we have looked at the bulk properties like
transverse momentum spectra, yields, average transverse momentum, and
various ratios corresponding to $\pi^{\pm}, K^{\pm}, p,$ and
$\bar{p}$. 
Table~\ref{tab:sum} gives a summary of the performance of various sets
in AMPT default and SM models for different observables presented. 
 This study also complements the $v_2$ comparison from AMPT with experimental
 data previously published at these energies.

The spectra comparisons in 0--5\% central collisions  
suggest that set A,
corresponding to 3 mb parton cross-section with $a=0.55$ and $b=0.15$ GeV$^{-2}$ as
Lund string fragmentation parameters in the AMPT model, 
seems to give a good description of
the spectra for all the considered particles at three energies in 0--5\% central collisions. 
The kaon spectrum in the SM AMPT model underpredicts the data by a factor 2
at 7.7 GeV, which suggests that the SM AMPT model is not well 
suited to study kaon (strangeness) production at lower energies. 

The comparisons of $dN/dy$ as a function of collision centrality suggest that
particle yields cannot be described simultaneously by the AMPT model
for all the particles at the three energies. 
Pion and proton $dN/dy$ are not described at 200
GeV while kaon $dN/dy$ is not described at 7.7 GeV at all the centralities.
The pion and proton $dN/dy$ at 7.7 GeV are described by set B parameters; pion and kaon $dN/dy$ at 27 GeV are described by default set C parameters while proton $dN/dy$ are described by SM set A parameters; and kaon $dN/dy$ at 200 GeV is described by default set A and SM set C parameters.
In general, for most cases, it is observed that set C parameters with the largest $a=2.2$ produce higher yields while set B parameters with the smallest $a=0.5$ produce smaller yields of particles.

Comparisons of \pTavg as a function of centrality suggest that
except for the pion \pTavg at 7.7 GeV (which is described at all centralities by
default set C parameters), the 
AMPT model does not explain the data at all centralities for the 
three energies. The pion \pTavg 
is not explained at all by the AMPT model at 200 GeV. For other energies
and particles, \pTavg as a function of
centrality is only explained partially by the model. 
For these cases,
the set A parameters, corresponding to 3 mb cross section with $a=0.55$
and $b=0.15$ GeV$^{-2}$, mostly work better than the other two. 

Comparisons of antiparticle-to-particle ratios as a function of
centrality suggest that
$\pi^{-}/\pi^{+}$ ratio is explained by the AMPT model at three energies
and all centralities
while $K^{-}/K^{+}$ and $\bar{p}/p$ ratios are explained at 27 and 200
GeV
but not at 7.7 GeV, where they are only partly
described.
The pion ratio does not distinguish between default and SM
AMPT models. The kaon ratio favors SM; similarly, the proton ratio
also favors the SM except at 200 GeV, where default AMPT works better.

Comparisons of mixed particle ratios as a function of centrality
suggest that the $K^{\pm}/\pi^{\pm}$ ratio at 7.7 GeV is not described
well by any of the models at all centralities, except at the
peripheral bins. This indicates that the production of kaons (strange
particles) is not explained by the AMPT model at lower energies.
 At 27 and 200 GeV, these ratios are described by default and/or SM
 set C parameters, i.e., with a 10 mb parton cross section and $a=2.2$, $b=0.5$ GeV$^{-2}$.  
The $p/\pi^{+}$ ratio is explained for all centralities at 7.7 and 27
GeV but for a few centralities at 200 GeV by the AMPT model. The model
explains the $\bar{p}/\pi^{-}$ ratio for all centralities at all
energies. It seems that both the default and SM models are favored by
the $p/\pi^{+}$ ratio at $\sqrt{s_{NN}} \le$ 27 GeV while the default
AMPT model is favored by the $\bar{p}/\pi^{-}$ ratio for all
centralities at the three energies.

The energy dependence of mixed particle ratios for 0--5\% central collisions is also studied. 
The $K^{\pm}/\pi^{\pm}$ ratio is not explained by the AMPT model at 7.7
GeV. At 27 and 200 GeV, the default AMPT model with all the three
sets 
and SM with set A parameters describe the data. 
Thus, we again observe that the strangeness (kaon)
production at $\sNN$ = 7.7 GeV is not explained by the AMPT model. 
In addition, the default model is in relatively closer agreement with
the data at lower
energies than the SM AMPT model. 
In the SM model, the parton density is quite dense as all the HIJING
strings are converted to partons. Thus, the model with SM is expected to work well in
high energy density regions. 
Our observation that it performs unsatisfactorily (and worse than
default AMPT) at the lower energy of 7.7 GeV 
but gives progressively better predictions at higher energies
supports the above statement. 
The energy dependence of $p/\pi^+$ ratio is better explained by
default set A parameters and by SM set A and B parameters. 
The energy dependence of the $\bar{p}/\pi^-$ ratio is not consistently
described by any 
parameter set of either the default or SM model.
However, the default
AMPT can describe the ratio at the three energies with different sets,
with the
set A parameters results being closest to the data. 
These results, and also those from the 
energy dependence 
of particle yields, \pTavg, and ratios 
in 0--5\% central collisions, 
suggest that 
the default AMPT with set A parameters (3 mb cross section, $a=0.55$,
and $b=0.15$ GeV$^{-2}$) is generally better than  the 
other sets and also better than the SM version with any set.
As mentioned earlier, by including
the finite nuclear thickness of incoming colliding nuclei into the SM
model might improve results at lower energies.
Nevertheless, these results complement and are also in contrast to the previously established picture
that elliptic flow $v_2$ is better described by the SM 
version. 
This also suggests that
some more work might be needed
to consistently describe all bulk
properties of the system formed in heavy-ion collisions at different
energies and centralities. The new quark coalescence  is one of the steps towards
that direction~\cite{He:2017tla}. 
The results from this study provide input in constraining the models in a better way and also understanding the particle production in heavy-ion collisions.

\section{Acknowledgements}
We thank Bedangadas Mohanty for suggestions and comments on the manuscript.
A.N. acknowledges support from the IAS-NASI-INSA academies as part of the Science Academies' Summer Research Fellowship Program and the Physics Department, Panjab University Chandigarh.
N.S. acknowledges the support of a DST-SERB Ramanujan Fellowship (D.O. No. SB/S2/RJN-084/2015). L.K. acknowledges the support of SERB Grant No. ECR/2016/000109.  
 
\bibliography{biblio}   

\end{document}